\documentclass[pra,aps,10pt,preprint,showkeys,preprintnumbers,superscriptaddress,floatfix]{revtex4-1}

\usepackage{graphicx}		
\usepackage{bm}				
\usepackage{amsmath}
\usepackage{color}

\begin{document}

\title{Magnetoelastic waves in thin films}

\author{Frederic Vanderveken}
\email{Frederic.Vanderveken@imec.be}
\affiliation{Imec, 3001 Leuven, Belgium}
\affiliation{KU Leuven, Departement Materiaalkunde, SIEM, 3001 Leuven, Belgium}

\author{Florin Ciubotaru}
\email{Florin.Ciubotaru@imec.be}
\affiliation{Imec, 3001 Leuven, Belgium}
\author{Christoph Adelmann}
\email{Christoph.Adelmann@imec.be}
\affiliation{Imec, 3001 Leuven, Belgium}

\date{\today}

\begin{abstract}
This paper discusses the physics of magnetoelasticity and magnetoelastic waves as well as their mathematical description. Magnetoelastic waves occur as a result of strong coupling between spin waves and elastic waves in magnetostrictive ferromagnetic media. In a first part, the basic behavior of spin waves is reviewed in both bulk ferromagnets as well as in thin films. Next, elastic waves are discussed with a focus on thin films. Finally, the interactions between the elastic and magnetic domains are described and it is shown how this results in the formation of magnetoelastic waves. Based on the description of bulk magnetoelastic waves, a theory for magnetoelastic waves in thin films is developed and their dispersion relations are derived. It is shown that the behavior strongly depends on the geometry of the system, especially on the polarization of spin and elastic waves and the direction of the magnetization of the magnetostrictive ferromagnet.
\end{abstract}


\maketitle

The coupling between different physical properties of a system is of great interest for transducer elements. In recent years, the field of spintronics, which includes applications of magnetism and magnetic materials in electronics, has gained enormous attention, which has led to the introduction of commercial magnetic memory technologies. However, the energy conversion from the electric to the magnetic domain and \emph{vice versa} still remains challenging. Current magnetic memories are based on spin-transfer or spin-orbit torques \cite{Diao07,Bapna18}, which depend on the current density in the device and require typically energies in the order of (10s of) fJ to switch the magnetization direction of a nanomagnet that is characterized by an intrinsic energy barrier in the order of aJ. While fJ energies are promising for nonvolatile memories, they are not competitive in other spintronic applications, such as spintronic logic circuits. Therefore, much research has been devoted to developing efficient transducers between electric and magnetic subsystems of spintronic devices. 

One of the most promising approaches to efficiently couple electric and magnetic properties in a spintronic system are magnetoelectric transducers. Magnetoelectric transducers consist of composite materials, which comprise piezoelectric and magnetostrictive components \cite{Polzikovaa16,Polzikova18,Cherepov14,Balinskiy18, Zhou14}. These piezoelectric-magnetostrictive bilayers or multilayers have been found to be particularly efficient. Applying a voltage to the piezoelectric layer(s) leads to the formation of strain in the compound. This strain introduces an effective magnetic anisotropy field in the magnetostrictive ferromagnetic component, leading to an effective coupling between voltage (electric field) and magnetization. The coupling is bidirectional since rotating the magnetization of a magnetostrictive layer also induces strain in the compound and consequently a polarization in the piezoelectric. Hence, such magnetoelectric schemes provide indirect coupling between electricity and magnetism mediated by elastodynamics. 

The coupling scheme of a magnetoelectric transducer can be split into two parts, (i) piezoelectric coupling between electric and elastic domains, and (ii) magnetostrictive coupling between elastic and magnetic domains. Here, we investigate the second part and focus especially on the behavior at GHz frequencies. In this frequency range, elastic waves (hypersound) interact with magnetic waves (spin waves), forming hybrid magnetoelastic waves under resonant conditions. The physics of the magnetoelastic resonance and the resulting magnetoelastic waves have been described in bulk systems decades ago \cite{Kittel58,Akhiezer59,Fedders74,Kobayashi73,Kobayashi733,Kamra14,Keller15,Tucker72,Gurevich96, Schlomann60}. However, modern applications of magnetoelastic waves in magnetoelectric and spintronic devices are based on nm-thick films. Currently, a comprehensive understanding of the properties of magnetoelastic waves in thin films is still lacking. This paper aims at filling this gap by developing a theory of magnetoelastic waves in thin films based on the established theories for both elastic and spin waves. 

The paper begins by introducing basic magnetic interactions and by reviewing the properties and dispersion relations of spin waves in bulk ferromagnets and thin films. In the second section, linear elasticity and elastic waves in thin films are discussed. In the third section, the magnetoelastic interactions are introduced and the properties of magnetoelastic waves in thin films are derived, including their dispersion relations. 

\section{Spin waves}
\label{sec:spin_waves}

Spin waves are defined as collective excitations of the magnetization in magnetic materials. The properties of spin waves are strongly affected by the geometry and the dominant interactions inside the material. Hence, the relevant magnetic interactions will be shortly introduced, followed by the derivation of the properties of spin waves in bulk and thin film ferromagnets using the plane wave method. 

\subsection{Spin waves in bulk ferromagnets}
\label{sec:spin waves bulk}

The magnetization dynamics in a ferromagnet can be described by the Landau–Lifshitz– Gilbert (LLG) equation \cite{Landau35,Gilbert04}
\begin{equation}
\label{eq:llg}
	\frac{d\mathbf{M}}{dt} = -\gamma \mu_0(\mathbf{M} \times \mathbf{H}_{\mathrm{eff}}) + \frac{\alpha}{M_\mathrm{s}} \left( \mathbf{M} \times \frac{d\mathbf{M}}{dt} \right)
\end{equation}

\noindent with $-\gamma$ the gyromagnetic ratio [s$^{-1}$T$^{-1}$], $\mu_0$ the vacuum permeability [TmA$^{-1}$], $\alpha$ the Gilbert damping constant, $M_\mathrm{s}$ the saturation magnetization [Am$^{-1}$], and $H_{\mathrm{eff}}$ the effective magnetic field [Am$^{-1}$]. The first term in the LLG equation describes the precession of the magnetization around the effective magnetic field. The second term in the LLG equation leads to the damping of the magnetization precession towards the direction of the effective magnetic field. 

There are different magnetic interactions such as the exchange interaction, the dipolar interaction, the Zeeman interaction, magnetocrystalline anisotropy fields, magnetoelastic fields, \emph{etc}. If the interaction energy density $\mathcal{E}(\mathbf{M})$ is known, it is possible to calculate a magnetic field that corresponds to this interaction. This field is given by
\begin{equation}
\label{eq:heff}
	\mathbf{H} = -\frac{1}{\mu_0} \frac{d\mathcal{E}(\mathbf{M})}{d\mathbf{M}} \,.
\end{equation}

\noindent The total effective magnetic field, which is governing the magnetization dynamics in the LLG equation, is given by the sum of all individual magnetic fields, including externally applied fields. Below, the dipolar and exchange interaction are explained in more detail since these give rise to spin waves. Full elaborated discussions of spin waves and their properties can be found in \cite{Gurevich96, Stancil09, Cottam94}.

The dipolar interaction describes the direct interaction between magnetic dipoles. Following Eq.~(\ref{eq:heff}), the interaction can be represented by a dipolar magnetic field. This field is found by solving Maxwell's equations. To describe spin waves at GHz frequencies, the magnetostatic approximation is valid since the wavelengths of such spin waves are several orders of magnitude shorter than those of electromagnetic waves in vacuum at the same frequency, \emph{i.e.} $k_0\ll k_\mathrm{sw}$ with $k_0$ the wavenumber of an electromagnetic wave in vacuum and $k_\mathrm{sw}$ the wavenumber of a spin wave. This approximation implies that the change in electric field over time, $\partial\mathbf{E}/{\partial t}$, has a negligible effect on the curl of the magnetic field, $\nabla \times \mathbf{H} = 0$. Assuming further that no free charges and no electrical currents are present inside the material, Maxwell's equations become
\begin{eqnarray}
\label{eq:maxwel1}
	\nabla \cdot \mathbf{E} &=& 0 \\
\label{eq:maxwel2}
	\nabla \cdot \mathbf{B} &=& 0 \\
\label{eq:maxwel3}
	\nabla \times \mathbf{E} &=& -\frac{\partial\mathbf{B}}{\partial t} \\
\label{eq:maxwel4}
	\nabla \times \mathbf{H} &=& 0 
\end{eqnarray}

\noindent with $\mathbf{B}=\mu_0(\mathbf{H}+\mathbf{M})$ the magnetic induction [T]. Hence, in the magnetostatic limit (as in the electrostatic limit), electric and magnetic fields are decoupled from each other. Equation (\ref{eq:maxwel4}) indicates that the curl of the magnetic field equals zero. This allows for the definition of a magnetic scalar potential $\phi$ as
\begin{equation}
\label{eq:Hdip}
	\mathbf{H}_{\mathrm{dip}} = - \nabla \phi \, .
\end{equation}
\noindent Using Eq. (\ref{eq:maxwel2}), the definitions of the magnetic scalar potential and the magnetic induction $\mathbf{B}$, one finds the magnetic Poisson relation
\begin{equation}
\label{eq:poisson}
	\nabla^2 \phi = \nabla \cdot \mathbf{M} \, .
\end{equation}
This shows that the divergence of the magnetization $\nabla \cdot \mathbf{M}$, also called the magnetic charge, acts as a source of the magnetic scalar potential and hence as a source for the dipolar field. Two types of magnetic charges can be identified: first, a surface charge, originating from surfaces between two materials with different magnetization magnitude or direction, and, second, a magnetic volume charge, originating from the change of the magnetization in the bulk of a ferromagnetic material. Both surface and volume magnetic charges generate dipolar fields. The field outside the magnetic material is called the stray field and the field inside the material is called the demagnetization field. 

By solving the magnetic Poisson equation (\ref{eq:poisson}) and using Eq. (\ref{eq:Hdip}), it is possible to derive a general expression for the demagnetization field given by \cite{Joseph65,Smith10}
\begin{equation}
\label{eq:Hdemag}
	\mathbf{H}_{\mathrm{demag}} = \frac{1}{4\pi} \int_{V'} \bar{D}(\mathbf{r}-\mathbf{r}')\mathbf{M}(\mathbf{r}') dV'
\end{equation}
\noindent with $V'$ the volume of the magnetic material and $\bar{D}(\mathbf{r}-\mathbf{r}')$ the tensorial magnetostatic Green's function given by \cite{Gurevich96}
\begin{equation}
	\bar{D}(\mathbf{r}-\mathbf{r}') = - \nabla_\mathbf{r} \nabla_\mathbf{r'} \frac{1}{|\mathbf{r}-\mathbf{r}'|}\,.
\end{equation} 
For uniform magnetization, the demagnetizing field is only generated by surface charges, and Eq. (\ref{eq:Hdemag}) reduces to
\begin{equation}
	\mathbf{H}_{\mathrm{demag}} = \frac{1}{4\pi} \mathbf{M} \int_{V'} \bar{D}(\mathbf{r}-\mathbf{r}') dV' = - \bar{N}(\mathbf{r}) \mathbf{M}
\end{equation}

\noindent with $\bar{N}(\mathbf{r})$ the demagnetization tensor, which only depends on the shape of the magnetic volume. The anisotropy introduced by the demagnetization field is thus often called the shape anisotropy. 

The second magnetic interaction that is necessary to describe spin waves is the exchange interaction between individual magnetic dipoles, which gives rise to ferromagnetic coupling below the Curie temperature \cite{Chikazumi86}. The exchange energy density is given by

\begin{equation}
\label{eq:Eex}
	\mathcal{E}_{\mathrm{ex}} = \frac{A_{\mathrm{ex}}}{M_{\mathrm{s}}^2} \left[(\nabla M_\mathrm{x})^2+(\nabla M_\mathrm{y})^2 + (\nabla M_\mathrm{z})^2\right]
\end{equation}

\noindent with $A_{\mathrm{ex}}$ the exchange stiffness constant [J/m]. Following Eq. (\ref{eq:heff}), the exchange field is

\begin{equation}
\label{eq:Hex}
	\mathbf{H}_{\mathrm{ex}} = \frac{2A_{\mathrm{ex}}}{\mu_0 M_\mathrm{s}^2} \Delta \mathbf{M} = l_{\mathrm{ex}}^2 \Delta\mathbf{M} \equiv \lambda_{\mathrm{ex}} \Delta \mathbf{M}
\end{equation}

\noindent with $\Delta$ the Laplace operator and $l_{\mathrm{ex}}$ the exchange length [m]. In ferromagnets, the exchange interaction tries to keep the individual magnetic moments parallel. The exchange length $l_{\mathrm{ex}}$ characterizes the competition between the dipolar and the exchange interaction \cite{Cottam94,Hillebrands02}. At length scales below the exchange length $l_{\mathrm{ex}}$, the exchange interaction is dominant and magnetic moments align parallel with each other. At length scales above the exchange length, the dipolar interaction is dominant, and it becomes possible for domains to form. Analogously, the properties of spin waves with short wavelengths are dominated by the exchange interaction, whereas the dipolar interaction strongly affects the properties of spin waves with large wavelengths. 

We consider now a ferromagnetic material with a static magnetic field $\mathbf{H}_{\mathrm{ext}}$ applied in the z-direction. In absence of any anisotropy, the external field forces the equilibrium magnetization along the z-direction. In such a system, stable wave-like excitations exist, which can be described by weak perturbations of the equilibrium magnetization. For a plane wave, the magnetization at a specific point in space and time can be written as

\begin{equation}
\label{eq:M_ansatz}
	\mathbf{M}(\mathbf{r},t) = \mathbf{M}_0 + \mathbf{m}(\mathbf{r},t) = \begin{bmatrix}
	0 \\ 0 \\ M_0
	\end{bmatrix} + \begin{bmatrix}
		m_\mathrm{x}  \\
		m_\mathrm{y}  \\
		0
	\end{bmatrix} e^{i(\omega t + \mathbf{k}\cdot \mathbf{r})} 
\end{equation}

\noindent with $\omega$ the angular frequency of the wave [rad s$^{-1}$], and $\mathbf{k}$ the wavevector with norm $||\mathbf{k}||=k=2\pi/\lambda$ [rad m$^{-1}$] and direction perpendicular to the phase front. For weak perturbations, \emph{i.e.} $||\mathbf{m}|| \ll M_0$, $\mathbf{m}(\mathbf{r},t)$ describes a wave-like perturbation which is called a spin wave.

In Eq. \ref{eq:M_ansatz}, the z-component of the dynamic magnetization, $m_\mathrm{z}$, is neglected. This approximation is only valid if the perturbations are weak. Since the angular momentum, \textit{i.e.} the norm of the magnetization vector, is conserved, the z-component is given by $m_\mathrm{z}^2 = M_0^2 - m_\mathrm{x}^2 - m_\mathrm{y}^2$. Therefore, the $m_\mathrm{z}$ component can be considered as a second order perturbation and is thus neglected in the remainder of this paper. 

For a uniform bulk material, the dipolar and exchange fields that correspond to the perturbed magnetization state can be found via Eqs. (\ref{eq:Hdemag}) and (\ref{eq:Hex}), respectively, and are given by \cite{Herring51}

\begin{equation}
	\mathbf{h}_\mathrm{dip}(\mathbf{r},t) = -\frac{\mathbf{k} \cdot \mathbf{m}(\mathbf{r},t)}{||\mathbf{k}||^2} \mathbf{k} = - \frac{1}{k^2}\begin{bmatrix}
	k_\mathrm{x}^2 & k_\mathrm{x}k_\mathrm{y} & 0 \\  k_\mathrm{x}k_\mathrm{y} & k_\mathrm{y}^2 & 0 \\ 0 & 0 & 0 
	\end{bmatrix} \mathbf{m}(\mathbf{r},t)
\end{equation}

\noindent and 

\begin{equation}
\mathbf{h}_{\mathrm{ex}}(\mathbf{r},t) = -\lambda_{\mathrm{ex}} k^2 \mathbf{m}(\mathbf{r},t)\, , 
\end{equation}
 
\noindent respectively. The wavevector $\mathbf{k}=[k_x,k_y,k_z]$ is determined by a single parameter $\theta$ because of the axial symmetry around the magnetization vector. It can thus be written as $\mathbf{k}=k[\sin(\theta),0,\cos(\theta)]$ with $\theta$ the angle between the magnetization and the propagation direction of the wave. With this substitution, the dipolar field can be simplified, leading to 
\begin{equation}
\mathbf{h}_{\mathrm{dip}}(\mathbf{r},t) = - \begin{bmatrix}
\sin^2(\theta) & 0 & 0 \\  0 & 0 & 0 \\ 0 & 0 & 0 
\end{bmatrix} \mathbf{m}(\mathbf{r},t) \, .
\end{equation}

The magnetization dynamics corresponding to the spin wave is found by solving the LLG equation (\ref{eq:llg}) including the perturbation $\mathbf{m}(\mathbf{r},t)$. Neglecting the damping term, we obtain
\begin{equation}
	\frac{d[\mathbf{M}_0 + \mathbf{m}(\mathbf{r},t)]}{dt} = -\gamma_0 [(\mathbf{M}_0 + \mathbf{m}(\mathbf{r},t)) \times (\mathbf{H}_{\mathrm{ext}}+ \mathbf{h}_{\mathrm{dip}}(\mathbf{r},t)+\mathbf{h}_{\mathrm{ex}}(\mathbf{r},t))]
\end{equation}
\noindent with $\gamma_0 = \gamma\mu_0$. Because the perturbation is assumed to be weak, terms quadratic in $\mathbf{m}$ can be neglected. By applying a temporal Fourier transformation, a linearized LLG equation is obtained, given by
\begin{equation}
\label{eq:llg_lin}
	i\omega \mathbf{m}(\mathbf{r},\omega) = -\gamma_0 [\mathbf{M}_0 \times (\mathbf{h}_{\mathrm{ex}}(\mathbf{r},\omega) +\mathbf{h}_{\mathrm{dip}}(\mathbf{r},\omega)) +  \mathbf{m}(\mathbf{r},\omega) \times \mathbf{H}_{\mathrm{ext}}] \, .
\end{equation}
\noindent Rearranging the terms and rewriting the system in matrix notation leads to
\begin{equation}
\label{eq:llg_lin_matrix}
	\begin{bmatrix}
	\omega_\mathrm{bx} & -i\omega \\
	i\omega & \omega_\mathrm{by}
	\end{bmatrix} \begin{bmatrix}
	m_\mathrm{x} \\ m_\mathrm{y}
	\end{bmatrix} = 0 
\end{equation}
\noindent with
\begin{eqnarray}
	\omega_\mathrm{bx} &=& \omega_0 + \omega_\mathrm{M} (\lambda_{\mathrm{ex}}k^2 + \sin^2(\theta)) \,, \\
	\omega_\mathrm{by} &=& \omega_0+\omega_\mathrm{M} \lambda_{\mathrm{ex}} k^2 \,,
\end{eqnarray}
\noindent $\omega_0=\gamma_0 H_{\mathrm{ext}}$, and $\omega_\mathrm{M}=\gamma_0 M_\mathrm{s}$. The parameters $\omega_\mathrm{bx}$ and $\omega_\mathrm{by}$ are related to the effective magnetic fields that interact with the x- and y-components of the dynamic magnetization, respectively. 

The properties of the stable perturbations of the magnetization, \emph{i.e.} the spin waves, can be extracted by analyzing the eigenvalues and corresponding eigenstates of Eq. (\ref{eq:llg_lin_matrix}). Equation (\ref{eq:llg_lin_matrix}) has nontrivial solutions only if its determinant is zero. This condition can be used to obtain the dispersion relations of the spin waves. Considering only positive frequencies, the spin wave angular frequency is given by

\begin{equation}
\label{eq:disp_spin_bulk}
\omega = \sqrt{\omega_\mathrm{bx}\omega_\mathrm{by}} = \sqrt{(\omega_0+\omega_\mathrm{M} \lambda_{\mathrm{ex}}k^2)[\omega_0+\omega_\mathrm{M}( \lambda_{\mathrm{ex}}k^2 + \sin^2(\theta))]} \, .
\end{equation}

\noindent This equation is the dispersion relation for spin waves in bulk ferromagnets. It is also called the Herring-Kittel equation. 

From Eq. \ref{eq:disp_spin_bulk}, it is noted that there is a minimum frequency (larger than zero) for which resonant magnetization dynamics can be obtained. Exciting a ferromagnet at frequencies below the spin wave resonance generates evanescent waves. If the excitation source is removed, these waves disappear after a certain time span (their lifetime) even in absence of intrinsic damping. Moreover, they do not propagate and thus do not contribute to steady state wave patterns at distances from the excitation source that are much larger than their wavelength. However, they are important to satisfy boundary conditions and in transient regimes. 

The eigenstates corresponding to the eigenvalues of the linearized LLG equation are 
\begin{equation}
\label{eq:eig_spin_bulk}
\mathbf{m}(k) = \frac{N}{\sqrt{\omega_\mathrm{bx}\omega_\mathrm{by}}} \begin{bmatrix}
i \omega_\mathrm{by} \\ \sqrt{\omega_\mathrm{bx}\omega_\mathrm{by}}
\end{bmatrix}
\end{equation}
\noindent with $N$ being a normalization constant. Note that $\omega_\mathrm{bx}$ and $\omega_\mathrm{by}$ both depend on $k$ and thus, via the dispersion relation, also on the frequency. The eigenstate indicates that the precession of the magnetization (the polarization of the wave) is always clockwise in the direction of propagation. Furthermore, the precession of the magnetization is generally elliptical with an ellipticity equal to $|m_\mathrm{x}|/|m_\mathrm{y}|$.

In the limit of small $k$, the exchange interaction can be neglected since $\lambda_\mathrm{ex} k^2 \ll 1$. Then, the dispersion relation becomes $ \omega = \sqrt{\omega_0(\omega_0+\omega_\mathrm{M}\sin^2(\theta))}$. This dispersion characterizes dipolar spin waves, which are degenerate and thus multiple spin waves with different wavelengths exist at the same frequency. For $\theta=0$, the effect of the dipolar self-interaction disappears, and one obtains $\omega=\omega_0$. In this case, the dynamic magnetization components only interact with the external field. For $\theta=\pi/2$, the interaction between the dynamic dipolar field and the spin wave is strongest. In this case, the dispersion relation is $\omega=\sqrt{\omega_0(\omega_0+\omega_M)}$. Therefore, the spin wave frequencies in the dipolar regime are limited to a specific interval
\begin{equation}
	\omega_0 \leq \omega \leq \sqrt{\omega_0(\omega_0+\omega_M)}\,.
\end{equation}
\noindent On the other hand, in the limit of large $k$, when $\lambda_\mathrm{ex} k^2 \gg 1$, one obtains a quadratic dispersion relation
\begin{equation}
\label{eq:disp_lim_exchange}
 \omega =  \omega_\mathrm{M} \lambda_{\mathrm{ex}} k^2 \,.
\end{equation}
\noindent This dispersion characterizes spin waves, for which the exchange interaction is dominant. It is worth noting that these exchange spin waves are isotropic with respect to the propagation direction. By contrast, dipolar spin waves are anisotropic because they depend on the propagation direction via the parameter $\theta$. 

\subsection{Spin waves in ferromagnetic thin films}
\label{sec:spin waves film}

In the previous section, the properties of spin waves in an infinite bulk medium were discussed. In this section, we introduce boundaries in the ferromagnetic medium and derive the properties of spin waves in ferromagnetic thin films. 

Consider an infinite magnetic thin film of thickness $d$ with its normal parallel to the y-direction. In the previous section, electrical currents were neglected in Maxwell's equations, which is only a good approximation for ferromagnetic insulators. However, for thin films, this approximation is valid even for conductors as long as the thickness of the film is sufficiently small with respect to the skin depth of the ferromagnet \cite{Tamaru11}. It should also be noted that in the derivations below, the dynamic magnetization and the fields are averaged over the thickness of the film and are thus uniform in the y-direction. This is a valid approximation when the wavelength is much larger than the thickness of the film, i.e. $kd \ll 1$. If this is not the case, it is possible for thickness modes to arise \cite{Gurevich96}, which will however not be considered here. 

The magnetization is again defined as in Eq. (\ref{eq:M_ansatz}) with the magnetization saturated in-plane along an external field $\mathbf{H}_\mathrm{ext}$ in the z-direction. The components of the dynamic magnetization are in the x- and y-direction and form a plane wave, the spin wave. The exchange field is not affected by the thin film boundaries and is given by Eq. (\ref{eq:Hex}). As indicated by Eq. (\ref{eq:poisson}), the boundaries generate magnetic surface charges and therefore act as a source of the dipolar field. Therefore, in contrast with the exchange field, the dipolar field is affected by the boundaries.

For a thin film, the dipolar field can be approximated as \cite{Damon61,Harte68,Kalinikos86}
\begin{eqnarray}
\mathbf{h}_{\mathrm{dip}}(\mathbf{r},t) &=& -[P\frac{\mathbf{k} \cdot \mathbf{m}}{||\mathbf{k}||^2} \mathbf{k} + (1-P)(\mathbf{n \cdot \mathbf{m}})\mathbf{n}] \\
 &=& - \begin{bmatrix}
P\sin^2(\theta) & 0 & P \sin(\theta)\cos(\theta) \\  0 & 1-P & 0 \\ P\sin(\theta)\cos(\theta) & 0 &  P\cos^2(\theta)
\end{bmatrix} \mathbf{m}(\mathbf{r},t)
\end{eqnarray}

\noindent with 
\begin{equation}
\label{eq:P}
	P = 1-\frac{1-e^{-kd}}{kd} \, ,
\end{equation}
\noindent $k^2 = k_\mathrm{x}^2+k_\mathrm{z}^2$, and $\theta$ the angle between the static magnetization $\mathbf{M}_0$ and wavevector $\mathbf{k}$. The linearized LLG equation (\ref{eq:llg_lin}) with the modified dipolar field can then be written as

\begin{equation}
\begin{bmatrix}
\omega_\mathrm{fx} & -i\omega \\
i\omega & \omega_\mathrm{fy}
\end{bmatrix} \begin{bmatrix}
m_\mathrm{x} \\ m_\mathrm{y}
\end{bmatrix} = 0 
\end{equation}
\noindent with
\begin{eqnarray}
\omega_\mathrm{fx} &=& \omega_0 + \omega_\mathrm{M} (\lambda_{\mathrm{ex}}k^2 + P\sin^2(\theta)) \\
\omega_\mathrm{fy} &=& \omega_0+\omega_\mathrm{M} (\lambda_{\mathrm{ex}} k^2 +1-P)\,.
\end{eqnarray}
\noindent Again, nontrivial solutions of this equation only exist when the determinant of the matrix is zero. This leads to the dispersion relation of spin waves in thin ferromagnetic films, given by \cite{Kalinikos86,Kalinikos81}
\begin{equation}
\label{eq:disp_spin_film}
\omega = \sqrt{\omega_\mathrm{fx}\omega_\mathrm{fy}} = \sqrt{(\omega_0 + \omega_\mathrm{M} \lambda_{\mathrm{ex}} k^2)(\omega_0 + \omega_\mathrm{M} \lambda_{\mathrm{ex}}k^2 + \omega_\mathrm{M} F_\mathrm{m})}
\end{equation}
\noindent with
\begin{equation}
F_\mathrm{m} = 1- P\cos^2(\theta) + \frac{\omega_\mathrm{M} P(1-P)\sin^2(\theta)}{\omega_0+\omega_\mathrm{M} \lambda_{\mathrm{ex}} k^2} \,.
\end{equation}

\noindent The corresponding eigenstate has the same form and properties as the eigenstate of spin waves in bulk ferromagnets and is given by 
\begin{equation}
\label{eq:eig_spin_film}
\mathbf{m}(k) = \frac{N}{\sqrt{\omega_\mathrm{fx}\omega_\mathrm{fy}}} \begin{bmatrix}
i \omega_\mathrm{fy} \\ \sqrt{\omega_\mathrm{fx}\omega_\mathrm{fy}}
\end{bmatrix}
\end{equation}
\noindent with $N$ a dimensionless normalization constant.

In the limit of large wavevectors (exchange limit, $\lambda_\mathrm{ex} k^2 \gg 1$), the dispersion relation reduces to Eq. (\ref{eq:disp_lim_exchange}) that was derived for bulk magnetic media. However, in the dipolar limit of small $k$-values ($\lambda_\mathrm{ex} k^2 \ll 1$), the dispersion relation differs from that in bulk ferromagnetic media. Again, two limiting cases can be found for $\theta=0$ and $\theta=\pi/2$.

\begin{figure}[tb]
\begin{center}
	\includegraphics[width=10cm]{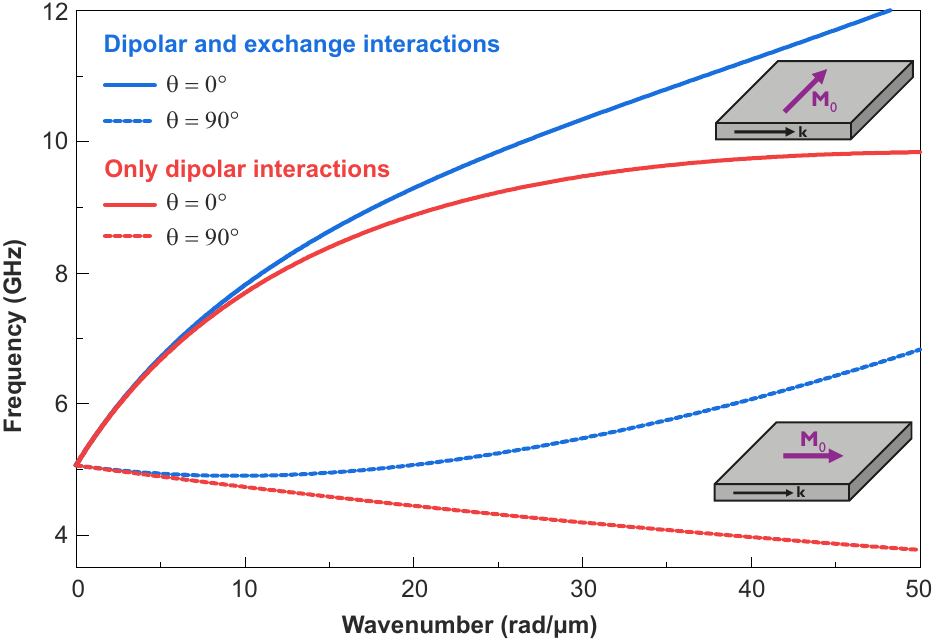}
\end{center}	
\caption{Spin wave dispersion according to Eq. (\ref{eq:disp_spin_film}) for a 30-nm-thick Ni film. Material parameters are $M_\mathrm{s}=480$ kA/m and  $A_{\mathrm{ex}}$=8 pJ/m, whereas the external magnetic field is $\mu_0H_\mathrm{ext} = 50$ mT. The solid red and blue lines correspond to dispersion relations of dipolar and dipolar--exchange surface spin waves, respectively. The dashed red and blue lines correspond to dispersion relations of dipolar and dipolar--exchange backward volume spin waves, respectively.}
	\label{fig:spin_wave_dispersions}      
\end{figure}

\noindent For $\lambda_\mathrm{ex} k^2 \ll 1$ and $\theta=0$, the dispersion relation becomes
\begin{equation}
\label{eq:disp_bvw}
\omega_{\mathrm{BVW}}^2 = \omega_0\left(\omega_0+ \omega_\mathrm{M} \frac{1-e^{-kd}}{kd}\right) \,.
\end{equation}
\noindent The propagation direction of these waves is parallel to the direction of the static equilibrium magnetization. Their dispersion relation is plotted in Fig. \ref{fig:spin_wave_dispersions} for a 30-nm-thick Ni film with $M_\mathrm{s}=480$~kA/m \cite{Cullity11}, $A_{\mathrm{ex}}=8$~pJ/m \cite{Wilts72}, and an external magnetic field of $\mu_0H_\mathrm{ext} = 50$ mT. According to the dispersion relation, the frequency decreases with increasing wavenumber and thus the group velocity, which is defined as $\mathbf{v}_\mathrm{g} =\partial \omega/\partial \mathbf{k}$, is negative. On the other hand, the phase velocity $\mathbf{v}_\mathrm{p}=\mathbf{k}\, \omega/k^2$, which describes the velocity and direction of the phase front, is positive. The energy flow of a wave is always parallel to the group velocity, and thus in this geometry, the energy flow and the group velocity are antiparallel to the wavevector and the phase velocity. For this reason, such waves are called backward volume waves (BVWs). As shown in Fig. \ref{fig:spin_wave_dispersions}, when the exchange interaction becomes non-negligible at larger wavevectors, dispersion relation shifts to higher frequencies. This effect increases for higher $k$-values, finally reaching the limiting case of exchange-dominated spin waves. 

In the dipolar limit ($\lambda_\mathrm{ex} k^2 \ll 1$), the dispersion relation for $\theta=\pi/2$ becomes
\begin{equation}
\label{eq:disp_surf_wave}
\omega_{\mathrm{SW}}^2 = \omega_0 (\omega_0+ \omega_\mathrm{M})+ \omega_\mathrm{M}^2 \left(1-\frac{1-e^{-kd}}{kd}\right)\frac{1-e^{-kd}}{kd} \,.
\end{equation}
\noindent These waves are called surface waves since their amplitude decays exponentially away from the surface. However, if the film is sufficiently thin, the magnetization can be considered uniform over the film thickness as mentioned earlier. The dispersion relations of the waves both in the dipolar approximation and when the dipolar and exchange interaction are simultaneously present are plotted in Fig. \ref{fig:spin_wave_dispersions}. The group velocity of this wave is positive and thus points in the same direction as the phase velocity. 

It should also be mentioned that spin waves are accompanied by an electric field and thus can be considered as an electromagnetic wave in a ferromagnetic medium. The dynamic electric field $\mathbf{e}$ is obtained from Maxwell's equations (\ref{eq:maxwel1}) and (\ref{eq:maxwel3}) and is determined by
\begin{equation}
	\nabla \cdot \mathbf{e} = 0
\end{equation}
\begin{equation}
\label{eq:electric_field}
	\nabla \times \mathbf{e} = -i \mu_0 \omega (\mathbf{h}_\mathrm{dip} + \mathbf{m}) = -i \mu_0 \omega \left( \bar{N}_\mathrm{dip}+ \bar{I} \right)  \mathbf{m} 
\end{equation}
\noindent with $\bar{I}$ the identity matrix. This indicates that both the dynamic dipolar field and the dynamic magnetization contribute to the generation of the dynamic electric field. In the magnetostatic approximation, $k_0\ll k$, the effect of the dipolar field is much smaller than that of the dynamic magnetization. Taking this into account and solving Maxwell's equations results in
\begin{equation}
	\mathbf{e} = -\frac{\mu_0 \omega}{k^2} \mathbf{k} \times \mathbf{m}
\end{equation}
\noindent for the electric field \cite{Stancil09}. At typical frequencies of spin waves in ferromagnetic media in the GHz range, the energy stored in the electric field is much smaller than the energy stored in the magnetic system \cite{Stancil09} and the magnetostatic waves can be considered as ``magnetization waves''. Note that this applies to spin waves in both the dipolar and exchange regime. In both cases, the wavelength of a spin wave is much shorter than the wavelength of an electromagnetic wave in vacuum at the same frequency and the magnetostatic approximation is thus valid. At higher frequencies reaching the THz regime, the magnetostatic approximation does no longer hold. The frequency condition for the magnetostatic approximation to be valid can be found by comparing the exchange and linear electromagnetic dispersion relations and becomes
\begin{equation}
\label{eq:magnetostatic}
	\omega \ll \frac{c^2}{\omega_\mathrm{M}\lambda_\mathrm{ex}} = \omega_\mathrm{crit}
\end{equation}
\noindent with $c$ is the speed of light in vacuum. For larger angular frequencies above $\omega_\mathrm{crit}$, the spin wave behaves similarly to classical electromagnetic waves with a considerable fraction of energy stored in the dynamic electric field.

\section{Elastic waves}
\label{sec:elastic waves}

In the previous section, the formation and properties of spin waves in ferromagnetic media, both in bulk materials and in thin films, has been discussed. In this section, we turn to the properties of wave-like oscillations of the displacement, \emph{i.e.} elastic waves. We start with a short derivation of the fundamental equations of linear elasticity and describe then the different types of elastic waves and their characteristics. 

\subsection{Elastodynamic equations of motion}
\label{sec:elastic waves equations}

The equation of motion for the displacement $\mathbf{u}$ is given by
\begin{equation}
\label{eq:elastodynamic}
	\rho \frac{d^2 \mathbf{u}}{dt^2} = \nabla \cdot \bar{\sigma} + \mathbf{f}_\mathrm{b} 
\end{equation}
\noindent with $\rho$ the mass density [kgm$^{-3}$], $\bar{\sigma}$ the two dimensional stress tensor with components $\sigma_\mathrm{ij}$ [Nm$^{-2}$] and $\mathbf{f}_\mathrm{b}$ the body forces acting on the material [Nm$^{-3}$]. For linear elastic materials, the stress tensor is related to the strain tensor via Hooke's law
\begin{equation}
\label{eq:hook}
	\bar{\sigma} = \bar{\bar{C}} : \bar{\varepsilon} \hspace{1 cm} \text{or} \hspace{1 cm} \sigma_\mathrm{ij} = \sum_{k=1}^{3} \sum_{l=1}^{3} C_\mathrm{ijkl} \varepsilon_\mathrm{kl}   \,.
\end{equation}
\noindent Here, $\bar{\bar{C}}$ is the fourth-order stiffness tensor and $\bar{\varepsilon}$ is the second-order strain tensor. Because of the symmetry of the stiffness tensor, it is possible to rewrite Hooke's law in reduced dimensionality \cite{Ewing57,Barber04} as
\begin{equation}
\label{eq:hook2}
\begin{bmatrix}
	\sigma_{11} \\ \sigma_{22} \\ \sigma_{33} \\ \sigma_{12} \\ \sigma_{13} \\ \sigma_{23}
\end{bmatrix} = \begin{bmatrix}
C_{11} & C_{12} & C_{13} &C_{14} &C_{15}  & C_{16} \\
       & C_{22} & C_{23} &C_{24} & C_{25} & C_{26} \\
       &        & C_{33} &C_{34} & C_{35} & C_{36} \\
       &        &        &C_{44} & C_{45} & C_{46} \\
 & \mathrm{symm.} &  &                         & C_{55} & C_{56} \\
 &  &  &  &                               & C_{66} \\
\end{bmatrix} \begin{bmatrix}
\varepsilon_{11} \\ \varepsilon_{22} \\ \varepsilon_{33} \\ 2\varepsilon_{12}\\ 2\varepsilon_{13} \\ 2\varepsilon_{23}
\end{bmatrix} \,.
\end{equation}
\noindent This is called the Voigt notation for Hooke's law.

Equation (\ref{eq:hook2}) indicates that a material with a nonsymmetric (\emph{e.g.} triclinic) crystal structure is described by 21 independent stiffness coefficients \cite{Graff75,Achenbach73}. In a crystal system with a certain symmetry, the number of independent stiffness constants can be greatly reduced. For example, only three independent stiffness constants are required to describe cubic crystal systems. The stiffness tensor then becomes
\begin{equation}
\label{eq:stiff_cubic}
	\bar{C}_\mathrm{cubic} = \begin{bmatrix}
	C_{11} & C_{12} & C_{12} &0 &0 & 0\\
	& C_{11} & C_{12} &0 & 0 & 0 \\
	&        & C_{11} &0 &0 & 0 \\
	&        &        &C_{44} & 0 & 0 \\
	& \mathrm{symm.} &  &  & C_{44} &0 \\
	&  &  &  &                               & C_{44} \\
	\end{bmatrix} \,.
\end{equation}
\noindent In this case, the residual anisotropy can be quantified by the Zener factor $A$, which is given by
\begin{equation}
A = \frac{2 C_{44}}{C_{11}-C_{12}}\,.
\end{equation} 
\noindent A Zener factor of 1 indicates fully isotropic elastic properties. In this isotropic limit, only two independent constants are required to describe the stiffness tensor. Different combinations of parameters can be used to represent the isotropic case, such as Young's modulus and the Poisson ratio, Young's modulus and the shear modulus, or the Lam\'e moduli. All descriptions are fully equivalent \cite{Achenbach73,Nayfeh95}.

For small displacements, the relation between the strain and displacement is given by \cite{Ewing57,Graff75}
\begin{equation}
\label{eq:strain_displacement}
\bar{\varepsilon} = \frac{1}{2} \left( \nabla \mathbf{u} + \left(\nabla \mathbf{u} \right) ^T - \nabla \mathbf{u} \left(\nabla \mathbf{u} \right) ^T \right) \approx \frac{1}{2} \left( \nabla \mathbf{u} + \left(\nabla \mathbf{u} \right) ^T \right) \,.
\end{equation}
\noindent Combining Eqs. (\ref{eq:elastodynamic}), (\ref{eq:stiff_cubic}), and (\ref{eq:strain_displacement}) results in the elastodynamic equations of motion with the displacement as the only variable. For a material with cubic anisotropy, the equations are given by
\begin{equation}
\begin{aligned}
\rho \frac{\partial^2 u_\mathrm{x}}{\partial t^2} &= C_{11} \frac{\partial^2 u_\mathrm{x}}{\partial x^2} + C_{44} \left( \frac{\partial^2 u_\mathrm{x}}{\partial y^2} +  \frac{\partial^2 u_\mathrm{x}}{\partial z^2} \right) + (C_{12} + C_{44}) \left( \frac{\partial^2 u_\mathrm{y}}{\partial x \partial y} + \frac{\partial^2 u_\mathrm{z}}{\partial x \partial z} \right) +f_\mathrm{x} \\
\rho \frac{\partial^2 u_\mathrm{y}}{\partial t^2} &= C_{11} \frac{\partial^2 u_\mathrm{y}}{\partial y^2} + C_{44} \left( \frac{\partial^2 u_\mathrm{y}}{\partial x^2} +  \frac{\partial^2 u_\mathrm{y}}{\partial z^2} \right) + (C_{12} + C_{44}) \left( \frac{\partial^2 u_\mathrm{x}}{\partial x \partial y} + \frac{\partial^2 u_\mathrm{z}}{\partial y \partial z} \right)+f_\mathrm{y} \\
\rho \frac{\partial^2 u_\mathrm{z}}{\partial t^2} &= C_{11} \frac{\partial^2 u_\mathrm{z}}{\partial z^2} + C_{44} \left( \frac{\partial^2 u_\mathrm{z}}{\partial x^2} +  \frac{\partial^2 u_\mathrm{z}}{\partial y^2} \right) + (C_{12} + C_{44}) \left( \frac{\partial^2 u_\mathrm{x}}{\partial x \partial z} + \frac{\partial^2 u_\mathrm{y}}{\partial y \partial z} \right)+f_\mathrm{z} \,.
\end{aligned}
\end{equation}
Remark that the above equations do not contain any damping terms and the system is assumed to be lossless. In practice, materials always possess some degree of viscoelasticity. In this case, the energy in the elastic wave is lost by different mechanisms such as phonon--phonon scattering due to the anharmonicity of the vibrational potential or the scattering of phonons by impurities. This can be taken into account by considering complex stiffness coefficients \cite{Achenbach73,Nayfeh95}. However, in the following, we will assume perfect elasticity without loss for simplicity. 

\subsection{Elastic waves in thin films}
\label{sec:elastic waves types}

In this section, we introduce the properties of elastic waves in an idealized thin film with free surfaces, which corresponds to an isolated thin film in vacuum. This model approximately represents also a thin film surrounded by materials with strongly different acoustic impedances, \emph{i.e.} a film with a strong acoustic impedance mismatch at its surfaces. For more realistic approaches, \emph{e.g.} a supported thin film on a substrate, appropriate stress and velocity boundary conditions need to be applied at the interfaces. In the next section, when the magnetoelastic interaction is included, it is demonstrated that the magnetization dynamics also generate elastic stresses, which further complicates the description at the boundaries. In such cases, an analytical treatment of the system is difficult and accurate studies will require numerical simulations, \emph{e.g.} by finite element methods. Nonetheless, the treatment of an idealized system presented here can provide analytical insights in the basic elastic (and magnetoelastic) behavior that can \emph{e.g.} be used to interpret numerical simulations of realistic systems.

For the case of thin films with free surface boundary conditions, the variation of the displacement along the thickness of the film is much smaller than the in-plane variation. Hence, the derivative of the displacement along the film surface normal can be neglected with respect to the derivatives in the in-plane directions, \emph{i.e.} $\partial u/ \partial y \ll \partial u/\partial x \,, \partial u/\partial z$. The elastodynamic equations of motion for a thin film with a surface normal in the y-direction are then given by
\begin{equation}
\begin{aligned}
\label{eq:elastodynamics_thin_film}
\rho \frac{\partial^2 u_\mathrm{x}}{\partial t^2} &= C_{11} \frac{\partial^2 u_\mathrm{x}}{\partial x^2} + C_{44}   \frac{\partial^2 u_\mathrm{x}}{\partial z^2}  + (C_{12} + C_{44})  \frac{\partial^2 u_\mathrm{z}}{\partial x \partial z}  \\
\rho \frac{\partial^2 u_\mathrm{y}}{\partial t^2} &= C_{44} \left( \frac{\partial^2 u_\mathrm{y}}{\partial x^2} + \frac{\partial^2 u_\mathrm{y}}{\partial z^2} \right) \\
\rho \frac{\partial^2 u_\mathrm{z}}{\partial t^2} &= C_{11} \frac{\partial^2 u_\mathrm{z}}{\partial z^2} + C_{44} \frac{\partial^2 u_\mathrm{z}}{\partial x^2}  + (C_{12} + C_{44}) \frac{\partial^2 u_\mathrm{x}}{\partial x \partial z}   \,.
\end{aligned}
\end{equation}
\noindent The above set of linear differential equations has wave-like solutions of the form \cite{Tucker72,Achenbach73}
\begin{equation}
\label{eq:U_ansatz}
	\mathbf{u}(\mathbf{r},t) = \begin{bmatrix}
	u_\mathrm{x} \\ u_\mathrm{y} \\ u_\mathrm{z}
	\end{bmatrix} e^{i(\omega t + \mathbf{k}\cdot \mathbf{r})}\,.
\end{equation}
\noindent To determine the dispersion relation of elastic waves in thin films, Eq. (\ref{eq:U_ansatz}) is substituted into the wave equations (\ref{eq:elastodynamics_thin_film}). Solving this system, applying both spatial and temporal Fourier transforms, and considering that the wavevector $\mathbf{k}$ points along the x-direction, results in
\begin{equation}
\label{eq:elastic_waves_matrix}
	\begin{bmatrix}
	\omega^2 - v_l^2k^2 & 0 & 0 \\
	0 & \omega^2 - v_\mathrm{t}^2k^2 & 0 \\
	0&0&\omega^2 - v_\mathrm{t}^2k^2
	\end{bmatrix} \begin{bmatrix}
	u_\mathrm{x} \\ u_\mathrm{y} \\ u_\mathrm{z}
	\end{bmatrix} = 0 
\end{equation}
\noindent with $v_l= \sqrt{C_{11}/\rho}$, $v_\mathrm{t} = \sqrt{C_{44}/\rho}$ and $\omega$ the angular frequency of the elastic wave. As a result, three independent elastic waves are found, which correspond to the three components of the displacement vector.

When only the $u_\mathrm{x}$ component is nonzero, longitudinal waves are formed since the oscillation is in the same direction as the wavevector. This wave is also frequently called a compressional or dilational wave. The dispersion relation, $\omega_l(k)$, of this wave is easily found from Eq. (\ref{eq:elastic_waves_matrix}) to be $\omega_l = v_lk$ \cite{Tucker72,Graff75,Achenbach73}. The dispersion relation is linear and thus the group velocity $v_l$ equals the phase velocity, independently of frequency. 

Waves with nonzero displacement components $u_\mathrm{y}$ and $u_\mathrm{z}$ oscillate perpendicular to the propagation direction. Therefore, these waves are transverse waves, also called shear or rotational waves. Their dispersion relation is also linear and equals $\omega_\mathrm{t} = v_\mathrm{t}k$ \cite{Tucker72,Graff75,Achenbach73}. The phase and group velocities are thus both equal to $v_\mathrm{t}$. It is further possible to classify these waves based on their polarization with respect to the film surface. The $u_\mathrm{y}$ component corresponds to shear vertical (SV) waves and the $u_\mathrm{z}$ component corresponds to shear horizontal (SH) waves. Furthermore, it is important to note that the velocity of the longitudinal wave is always larger than the velocity of the shear waves because $C_{11}> C_{44}$ \cite{Graff75,Achenbach73}.

The energy of elastic waves oscillates between the elastic potential energy and the kinetic energy. The elastic energy density is given by \cite{Tucker72,Graff75,Achenbach73}
\begin{equation}
\label{eq:elastic_energy}
\mathcal{E}_\mathrm{el} = \frac{1}{2} \bar{\sigma} :  \bar{\varepsilon} = \frac{1}{2} C_{ijkl} \varepsilon_{ij} \varepsilon_{kl} = \frac{1}{2} \sum_{i=1}^3 \sum_{j=1}^3 \sum_{k=1}^3 \sum_{l=1}^3 C_{ijkl} \varepsilon_{ij} \varepsilon_{kl}
\end{equation}
\noindent or in Voigt notation
\begin{equation}
\mathcal{E}_\mathrm{el} = \frac{1}{2} \bar{\sigma} :  \bar{\varepsilon} = \frac{1}{2} C_{ij} \varepsilon_{j} \varepsilon_{i}\,.
\end{equation}
\noindent The kinetic energy density is given by 
\begin{equation}
\label{eq:kinetic_energy}
\mathcal{E}_\mathrm{kin} = \frac{\rho||\mathbf{v}||^2}{2 } \hspace{15pt} \text{with} \hspace{15pt}  \mathbf{v} = \frac{\partial \mathbf{u}}{\partial t}\,.
\end{equation}
\noindent Hence, for an elastic wave the total energy is $\mathcal{E}_\mathrm{tot}=\mathcal{E}_\mathrm{el}+\mathcal{E}_\mathrm{kin}$ and $\mathcal{E}_\mathrm{el}=\mathcal{E}_\mathrm{kin}$.

\section{Magnetoelastic waves}
\label{sec:magnetoelastic waves}

In the two previous sections, magnetic and elastic waves in thin films were studied. Below, we now consider interaction terms between the magnetic and elastic domains, which couple magnetic and elastic waves. The section is divided into two parts. In the first part, the magnetoelastic interactions are described. In the second part, the properties of coupled magnetoelastic waves are derived. Magnetoelastic waves have been thoroughly studied in bulk materials \cite{Kittel58,Akhiezer59,Fedders74} or at free surfaces \cite{Ganguly76,Dreher12,Thevenard14}. By contrast, the properties of magnetoelastic waves in thin films has received little attention so far. Therefore, the focus in this paper will be on magnetoelastic waves in thin films, taking into account both exchange and dynamic dipolar magnetic fields.

\subsection{Magnetoelastic interactions}

The magnetoelastic interaction can be separated in two different effects. First, the influence of the direction of the magnetization on the internal strain in a ferromagnet, called the magnetostrictive effect. Second, the effect of strain on the magnetization state, called the Villari effect. If both effects are considered simultaneously, one speaks about magnetoelasticity.

\subsubsection{Magnetostriction}

Magnetostriction describes how the magnetization affects the elastic behavior of a material. Therefore, in a magnetostrictive material, different magnetization states result in different strain states. For a material with cubic symmetry, the magnetoelastic energy density is given by \cite{Kittel58}

\begin{equation}
\label{eq:Emel}
\begin{gathered}
\mathcal{E}_\mathrm{mel} = 
\frac{B_1}{M_\mathrm{s}^2}\left(\varepsilon_{\mathrm{xx}}\left(M_\mathrm{x}^2-\frac{1}{3}\right) + \varepsilon_{\mathrm{yy}}\left(M_\mathrm{y}^2-\frac{1}{3}\right) + \varepsilon_{\mathrm{zz}}\left(M_\mathrm{z}^2-\frac{1}{3}\right)\right) \\ 
+ \frac{2B_2}{M_\mathrm{s}^2} \left(\varepsilon_{\mathrm{xy}}M_\mathrm{x}M_\mathrm{y} + \varepsilon_{\mathrm{yz}}M_\mathrm{y}M_\mathrm{z} + \varepsilon_{\mathrm{zx}}M_\mathrm{x}M_\mathrm{z}\right)
\end{gathered}
\end{equation}

\noindent with $B_1$ and $B_2$ the linear isotropic and anisotropic magnetoelastic coupling constants, respectively [Jm$^{-3}$]. It is worth noting that the magnitude of the saturation magnetization has no influence on the magnetoelastic energy or strain state, which are rather determined by the orientation of the magnetization vector. The magnetization orientation is defined by the vector

\begin{equation}
\label{eq:def_alpha}
\boldsymbol{\zeta} = \begin{bmatrix}
\zeta_\mathrm{x} \\ \zeta_\mathrm{y} \\ \zeta_\mathrm{z}
\end{bmatrix} = \frac{1}{M_\mathrm{s}}\begin{bmatrix}
M_\mathrm{x} \\M_\mathrm{y} \\ M_\mathrm{z} 
\end{bmatrix}\,.
\end{equation}

\noindent Substituting this into Eq. (\ref{eq:Emel}) leads to

\begin{equation}
\begin{gathered}
\mathcal{E}_\mathrm{mel} = 
B_1\left(\varepsilon_{\mathrm{xx}}(\zeta_\mathrm{x}^2-\frac{1}{3}) + \varepsilon_{\mathrm{yy}}(\zeta_\mathrm{y}^2-\frac{1}{3}) + \varepsilon_{\mathrm{zz}}(\zeta_\mathrm{z}^2-\frac{1}{3})\right) \\ 
+ 2B_2 \left(\varepsilon_{\mathrm{xy}}\zeta_\mathrm{x}\zeta_\mathrm{y} + \varepsilon_{\mathrm{yz}}\zeta_\mathrm{y}\zeta_\mathrm{z} + \varepsilon_{\mathrm{zx}}\zeta_\mathrm{x}\zeta_\mathrm{z}\right)\,.
\end{gathered}
\end{equation}

\noindent Based on this expression for the magnetoelastic energy density, it is possible to calculate the magnetostrictive body force, which is given by

\begin{equation}
\label{eq:force_density1}
\mathbf{f}_\mathrm{mel} = \nabla \cdot \bar{\sigma}_\mathrm{mel} = \nabla \cdot \left( \frac{d\mathcal{E}_\mathrm{mel} }{d\varepsilon_\mathrm{ij}}\right) \, ,
\end{equation}

\noindent leading to

\begin{equation}
\label{eq:force_density2}
\renewcommand\arraystretch{1.8}  
\mathbf{f}_\mathrm{mel} = 2 B_1 \begin{bmatrix}
 \zeta_\mathrm{x} \frac{\partial \zeta_\mathrm{x}}{\partial \mathrm{x}} \\ \zeta_\mathrm{y} \frac{\partial \zeta_\mathrm{y}}{\partial y} \\ \zeta_\mathrm{z} \frac{\partial \zeta_\mathrm{z}}{\partial z} \end{bmatrix} + B_2 \begin{bmatrix}
  \zeta_\mathrm{x}\left( \frac{\partial \zeta_\mathrm{y}}{\partial \mathrm{y}} + \frac{\partial \zeta_\mathrm{z}}{\partial \mathrm{z}} \right)  + \zeta_\mathrm{y} \frac{\partial \zeta_\mathrm{x}}{\partial y} + \zeta_\mathrm{z} \frac{\partial \zeta_\mathrm{x}}{\partial z} \\
  \zeta_\mathrm{y}\left( \frac{\partial \zeta_\mathrm{x}}{\partial x} + \frac{\partial \zeta_\mathrm{z}}{\partial z} \right) + \zeta_\mathrm{x} \frac{\partial \zeta_\mathrm{y}}{\partial x} + \zeta_\mathrm{z} \frac{\partial \zeta_\mathrm{y}}{\partial z}  \\
 \zeta_\mathrm{z}\left( \frac{\partial \zeta_\mathrm{x}}{\partial x} + \frac{\partial \zeta_\mathrm{y}}{\partial y} \right)  + \zeta_\mathrm{x} \frac{\partial \zeta_\mathrm{z}}{\partial x} + \zeta_\mathrm{y} \frac{\partial \zeta_\mathrm{z}}{\partial y} 
\end{bmatrix}\,.
\end{equation}

\noindent There are three important parameters which determine the strength of the magnetostrictive body force: the magnetoelastic coupling constants, the magnetization orientation, and the gradient of the magnetization orientation.

The magnetostrictive body force affects the elastodynamics in the thin film and thus needs to be added to the elastodynamic equation (\ref{eq:elastodynamic}). This allows for the analytical description of the influence of the magnetization (direction) on the elastodynamics and the properties of the (magneto-)elastic waves. 

Another important quantity is the magnetostrictive strain which is the additional strain originating from the magnetostrictive effect. For a material with cubic anisotropy, the magnetostrictive strain is given by \cite{Chikazumi86,Tremolet93,Handley00} 

\begin{equation}
\label{eq:magnetostrictive_strain}
\bar{\varepsilon}_\mathrm{{mel}} = \frac{3}{2}\begin{bmatrix}
\lambda_{100}(\zeta_\mathrm{x}^2 -\frac{1}{3}) & \lambda_{111} \zeta_\mathrm{x} \zeta_\mathrm{y} & \lambda_{111} \zeta_\mathrm{x} \zeta_\mathrm{z} \\
\lambda_{111} \zeta_\mathrm{y} \zeta_\mathrm{x} & \lambda_{100}(\zeta_\mathrm{y}^2 -\frac{1}{3}) & \lambda_{111} \zeta_\mathrm{y} \zeta_\mathrm{z} \\
\lambda_{111} \zeta_\mathrm{z} \zeta_\mathrm{x} & \lambda_{111} \zeta_\mathrm{z} \zeta_\mathrm{y}& \lambda_{100}(\zeta_\mathrm{z}^2 -\frac{1}{3})
\end{bmatrix} \,.
\end{equation}

\noindent with $\lambda_{100}$ and $\lambda_{111}$ being the magnetostriction coefficients. $\lambda_{100}$ and $\lambda_{111}$ represent the maximum magnetostrictive strain for a fully-saturated magnetization along the $\langle 100\rangle$ or $\langle 111\rangle$ crystallographic directions, respectively. 

The magnetostriction coefficients are related to the magnetoelastic coupling constants by 

\begin{equation}
\lambda_{100} = \frac{2}{3} \frac{B_1}{C_{12}-C_{11}} \,, \hspace{1 cm} \lambda_{111} = -\frac{B_2}{3C_{44}} \,.
\end{equation}

\noindent Hence, it is also possible to express the magnetostrictive body force as a function of the magnetostrictive strain via

\begin{equation}
\label{eq:Hmel_body_force}
	\mathbf{f}_\mathrm{mel} = \nabla \cdot \bar{\sigma}_\mathrm{mel} = \nabla \cdot \left( \bar{\bar{C}}: \bar{\varepsilon}_\mathrm{mel}\right)  \,.
\end{equation}

\noindent Note that for an isotropic material, $\lambda_{100}=\lambda_{111} = \lambda_\mathrm{eq}$ and $B_1=B_2=B$. 

\subsubsection{The Villari effect}

The Villari effect describes how elastic strain affects the magnetization state. It is also called the inverse magnetostrictive effect. 

Strain in a magnetostrictive material results in an effective magnetoelastic field which can be derived from Eqs. (\ref{eq:heff}) and (\ref{eq:Emel}). For a material with cubic anisotropy, the magnetoelastic effective field is

\begin{equation}
\label{eq:Hmel}
\mathbf{H}_{\mathrm{mel}} = -\frac{1}{\mu_0} \frac{d\mathcal{E}_\mathrm{mel}}{d\mathbf{M}} = -\frac{2}{\mu_0M_\mathrm{s}}  \begin{bmatrix}
B_1\varepsilon_{\mathrm{xx}}\zeta_\mathrm{x} + B_2(\varepsilon_{\mathrm{xy}}\zeta_\mathrm{y}+\varepsilon_{zx}\zeta_\mathrm{z}) \\
B_1\varepsilon_{\mathrm{yy}}\zeta_\mathrm{y} + B_2(\varepsilon_{\mathrm{xy}}\zeta_\mathrm{x}+\varepsilon_{yz}\zeta_\mathrm{z}) \\
B_1\varepsilon_{\mathrm{zz}}\zeta_\mathrm{z} + B_2(\varepsilon_{\mathrm{zx}}\zeta_\mathrm{x}+\varepsilon_{yz}\zeta_\mathrm{y}) 
\end{bmatrix}
\end{equation}

\noindent with $\zeta_i$ the normalized magnetization components, as defined by Eq. (\ref{eq:def_alpha}). The resulting magnetization dynamics are described by the LLG equation (\ref{eq:llg}), including the above magnetoelastic field in $\mathbf{H}_{\mathrm{eff}}$.

\subsection{Magnetoelastic waves in thin films}

When the magnetoelastic interaction terms $\mathbf{f}_{\mathrm{mel}}$ and $\mathbf{H}_\mathrm{mel}$ are combined with the magnetodynamic equation (\ref{eq:llg}) and the elastodynamic equation (\ref{eq:elastodynamics_thin_film}), a set of coupled differential equations is obtained. Formally, these differential equations are nonlinear because the terms originating from the magnetoelastic interaction show a quadratic dependence on the magnetization and the displacement. Therefore, the magnetoelastic effect formally results in a nonlinear interaction. However, when the dynamic components are assumed to be weak, the differential equations can be linearized. In this case, wave-like solutions for the magnetization and the displacement, given by Eqs. (\ref{eq:M_ansatz}) and (\ref{eq:U_ansatz}), exist for the coupled set of equations. These solutions correspond to magnetoelastic waves. However, it is important to keep in mind that for large dynamic components, a system of nonlinear differential equations has to be solved, including nonlinear magnetoelastic interaction effects. 

To reduce the complexity of the calculations, a homogeneous and isotropic material is assumed. The geometry of the structure remains the same as in the previous sections with the film in the xz-plane and the y-direction normal to the film surface. The static magnetization and the static external field are chosen along the z-direction, as in Section \ref{sec:spin waves film}. Then, substituting the wave-like ansatz into the equations of motion, applying both spatial and temporal Fourier transforms, and neglecting terms quadratic in $\mathbf{m}$ or $\mathbf{u}$, leads to the following linearized system of equations:

\begin{equation}
\label{eq:system_of_melwaves}
\begin{aligned}
-\rho \omega^2 u_\mathrm{x} &= -C_{11} k_\mathrm{x} u_\mathrm{x} - C_{44}   k_\mathrm{z} u_\mathrm{x} -  (C_{12} + C_{44})  k_\mathrm{x} k_\mathrm{z} u_\mathrm{z} + \frac{B_2}{M_\mathrm{s}} ik_\mathrm{z}m_\mathrm{x} \\
-\rho \omega^2 u_\mathrm{y} &= -C_{44} \left( k_\mathrm{x} u_\mathrm{y}+ k_\mathrm{z} u_\mathrm{y} \right) + \frac{B_2}{M_\mathrm{s}} ik_\mathrm{z}m_\mathrm{y}\\
-\rho \omega^2 u_\mathrm{z} &= -C_{11} k_\mathrm{z} u_\mathrm{z} - C_{44} k_\mathrm{x} u_\mathrm{z}  - (C_{12} + C_{44}) k_\mathrm{x} k_\mathrm{z} u_\mathrm{x} + \frac{B_2}{M_\mathrm{s}}  ik_\mathrm{x}m_\mathrm{x} \\
i\omega m_\mathrm{x} &= -\omega_\mathrm{fy}m_\mathrm{y} - \gamma B_2 ik_\mathrm{z} u_\mathrm{y} \\
i\omega m_\mathrm{y} &= \omega_\mathrm{fx}m_\mathrm{x} + \gamma B_2 i \left(k_\mathrm{z} u_\mathrm{x} + k_\mathrm{y} u_\mathrm{z} \right)\,.
\end{aligned}
\end{equation}

\noindent Assuming that the elastic properties of the thin film are isotropic, the $C_{12}$ stiffness constant can be replaced by $C_{12} = C_{11}-2C_{44}$. Moreover, as discussed above, two types of elastic waves exist in in an isotropic material, \emph{i.e.} longitudinal and transverse waves. Therefore, it is convenient to define new displacement variables parallel ($u_l$) and perpendicular ($u_\mathrm{t}$) to the propagation direction so that

\begin{equation}
	u_\mathrm{x} = u_l\sin(\theta)+u_\mathrm{t}\cos(\theta) \,, \hspace{1 cm} u_\mathrm{z} = u_l\cos(\theta)-u_\mathrm{t}\sin(\theta)\,.
\end{equation}

\noindent Here, $\theta$ is the angle between the static magnetization $\mathbf{M}_0$ and the propagation direction $\mathbf{k}$. Substituting these redefined displacement components into the dynamic equation of motion together with $k_\mathrm{x} = k\cos(\theta)$ and $k_\mathrm{y} = k\sin(\theta)$ results in

\begin{equation}
\label{eq:system_of_melwaves2}
\begin{aligned}
(\omega^2-\omega_l^2)\sin(\theta) u_l  +  (\omega^2-\omega_\mathrm{H})\cos(\theta) u_\mathrm{t}  +  \frac{iBk\cos(\theta)}{\rho M_s} m_\mathrm{x}  & =  0\\
(\omega^2-\omega_\mathrm{V}^2) u_\mathrm{y} + \frac{iBk\cos(\theta)}{\rho M_s} m_\mathrm{y} & = 0  \\
(\omega^2-\omega_l^2)\cos(\theta) u_l  -  (\omega^2-\omega_\mathrm{H}^2)\sin(\theta) u_\mathrm{t}  +  \frac{iBk\sin(\theta)}{\rho M_s} m_\mathrm{x}  & =  0 \\
i\gamma Bk\cos(\theta)  u_\mathrm{y} + i\omega m_\mathrm{x} + \omega_\mathrm{fy}  m_\mathrm{y} & = 0 \\
i\gamma Bk\sin(2\theta) u_l + i\gamma Bk\cos(2\theta) u_\mathrm{t} + \omega_\mathrm{fx}  m_\mathrm{x}  -i\omega  m_\mathrm{y} & =0
\end{aligned}
\end{equation}

\noindent with $\omega$ the angular frequency of the magnetoelastic wave, $\omega_l=v_lk = \sqrt{\frac{C_{11}}{\rho}} k$ the dispersion relation of longitudinal elastic waves, $\omega_\mathrm{H}=v_\mathrm{t} k = \sqrt{\frac{C_{44}}{\rho}} k$ the dispersion relation of horizontally-polarized (in-plane) transverse elastic waves, and $\omega_\mathrm{V}=\omega_\mathrm{H}$ the dispersion relation of vertically-polarized (out-of-plane) transverse elastic waves. Here, the distinction between $\omega_\mathrm{V}$ and $\omega_\mathrm{H}$ is made to keep track of the origin of different terms in the equations of motion. In the following, different cases and geometries of magnetoelastic wave solutions of the coupled equations of motion are discussed. 

\subsubsection{Wave propagation perpendicular to the magnetization}

We first consider the case, in which the wave propagation direction is perpendicular to the static equilibrium magnetization, \emph{i.e.} $\theta=\pi/2$. In this geometry, Eq. (\ref{eq:system_of_melwaves2}) indicates that the magnetoelastic body force only acts on the $u_\mathrm{t}$ component of the displacement. Conversely, only the displacement component $u_\mathrm{t}$ generates a magnetoelastic field, which interacts with magnetic system. Hence, only the in-plane transverse elastic wave couples to surface spin waves and \emph{vice versa}. This means that the longitudinal and out-of-plane transverse elastic waves are independent from the magnetic system in a first order approximation and their dispersion relations remain unchanged, \emph{i.e.} $\omega_l = v_lk$ and $\omega_\mathrm{V}= v_t k$, respectively, as described in Section \ref{sec:elastic waves}.

Eliminating all uncoupled equations and using $\theta=\pi/2$ in Eq. (\ref{eq:system_of_melwaves2}), the system becomes

\begin{equation}
	\label{eq:stelsel_mel_de}
	\begin{bmatrix}
	 \omega^2 - \omega_\mathrm{H}^2  & \frac{iBk}{\rho M_\mathrm{s}} & 0 \\
	 -i\gamma Bk & \omega_\mathrm{fx} & -i \omega \\
	0 & i \omega & \omega_\mathrm{fy} 	
	\end{bmatrix} \begin{bmatrix}
	u_\mathrm{t} \\ m_\mathrm{x} \\ m_\mathrm{y}
	\end{bmatrix} = 0 
\end{equation}

\noindent with $\omega$ the angular frequency of the magnetoelastic wave and $\omega_\mathrm{H} = v_\mathrm{t} k$ the resonance frequency of the uncoupled horizontally-polarized transverse elastic wave. Note that in this geometry, the transverse displacement component is fully aligned in the z-direction, \emph{i.e.} $u_\mathrm{t} = u_\mathrm{z}$. To obtain nontrivial solutions, the determinant of the linear system must vanish and thus

\begin{equation}
\label{eq:disp_mel_de1}
	(\omega^2 - \omega_\mathrm{H}^2)(\omega^2-\omega_\mathrm{fm}^2)- Jk^2 \omega_\mathrm{fy} = 0
\end{equation}

\noindent with

\begin{equation}
J = \frac{\gamma B^2}{\rho M_\mathrm{s}} \\
\end{equation}

\noindent and the uncoupled spin wave resonance frequency $\omega_\mathrm{fm} = \sqrt{\omega_\mathrm{fx}\omega_\mathrm{fy}}$. Equation (\ref{eq:disp_mel_de1}) has the general form of a dispersion relation of two interacting waves. The first wave is a transverse elastic wave characterized by $\omega^2 - \omega_\mathrm{H}^2 = 0$ and the second wave is a spin wave characterized by $\omega^2-\omega_\mathrm{fm}^2=0$. The interaction between these two waves is quantified by $Jk^2\omega_\mathrm{fy}$. As expected, setting the magnetoelastic coupling constant $B$ to zero leads to the original dispersion relations of uncoupled elastic and magnetic waves.

Equation (\ref{eq:disp_mel_de1}) has two physically-meaningful solutions for $\omega$, which are given by

\begin{equation}
\label{eq:disp_mel_de2}
\omega^2_{\pm} = \frac{\omega_\mathrm{H}^2 + \omega_\mathrm{fm}^2}{2} \pm \sqrt{\left(\frac{\omega_\mathrm{fm}^2 - \omega_\mathrm{H}^2}{2}\right)^2  + J k^2 \omega_\mathrm{fy}} \,.
\end{equation}

\noindent These two solutions represent the dispersion relations of the resulting magnetoelastic waves. 

\begin{figure}[tb]
\begin{center}
	\includegraphics[width=10cm]{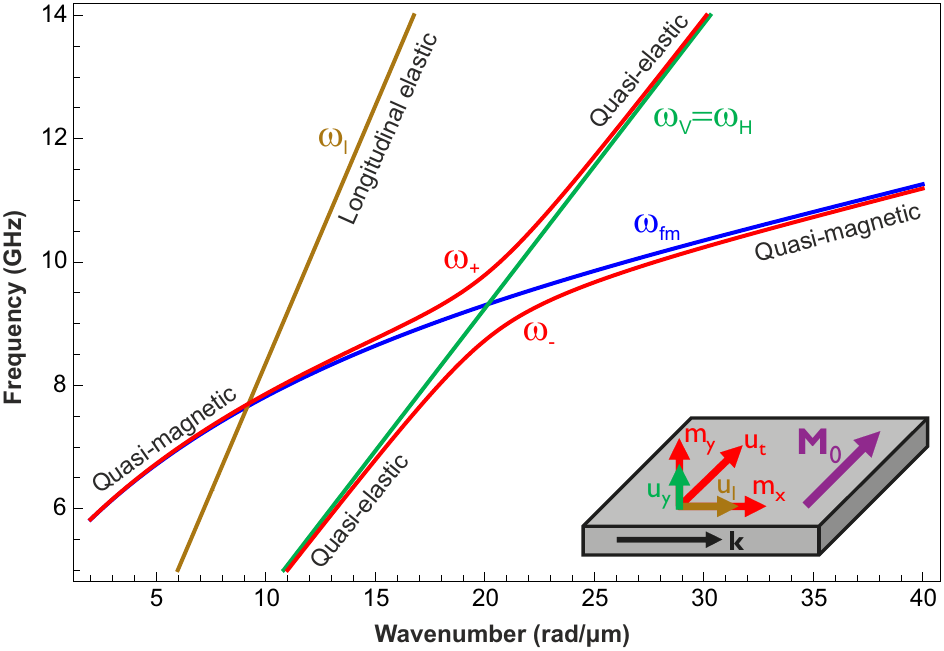}
\end{center}
\caption{Magnetoelastic wave dispersion relations according to Eq. (\ref{eq:disp_mel_de1}) for a 30-nm-thick Ni film and propagation direction perpendicular to the magnetization (red lines). The external magnetic field is $\mu_0H_\mathrm{ext} = 50$ mT. For comparison, the dispersion relations of longitudinal and transverse elastic waves (brown and green lines, respectively) as well as uncoupled spin waves (blue line) are also shown.}
	\label{fig:de_wave_dispersions}     
\end{figure}

The dispersion relations of the magnetoelastic and uncoupled elastic waves are plotted in Fig. \ref{fig:de_wave_dispersions} for a 30-nm-thick Ni film. The magnetic parameters are the same as used in Fig. \ref{fig:spin_wave_dispersions}. The magnetoelastic coupling constant is $B=10$~MJ/m$^3$ \cite{Walowski08, Sander99}, the stiffness constants are $C_{11}=245$~GPa and $C_{44}=75$~GPa \cite{Yamamoto50}, and the mass density is $\rho=8900$~kg/m$^3$ \cite{Mills99}. The two linear dispersion relations correspond to the uncoupled elastic waves. By contrast, the two red curves represent the different branches of the dispersion relation of the magnetoelastic waves with $m_\mathrm{x}$, $m_\mathrm{y}$, and $u_\mathrm{t}$ components. 

Figure \ref{fig:de_wave_dispersions} clearly shows that the two branches of the magnetoelastic wave dispersion relation do not cross each other. If the transverse elastic waves and the spin waves were not interacting, their dispersion relations would intersect. However, due to the magnetoelastic interaction, this crossing is avoided, leading to a gap between the two curves. This so-called anticrossing behavior of the dispersion relations is a typical characteristic of interacting waves \cite{Tucker72,Gurevich96}. 

The gap formation is also visible in the dispersion relations of Eq. (\ref{eq:disp_mel_de2}) itself. At the point where the noninteracting waves would be degenerate, the term $\omega_\mathrm{fm}^2 - \omega_\mathrm{H}^2$ vanishes. Then, the interaction coefficient $ Jk^2\omega_\mathrm{fy}$ has a strong influence on the dispersion. For $ Jk^2\omega_\mathrm{y} \gtrsim \omega_\mathrm{fm}^2 - \omega_\mathrm{H}^2$, the interaction between the magnetic and elastic system is strong, leading to the formation of coupled magnetoelastic waves. This results in the anticrossing with a frequency gap that quantifies the strength of the interaction and is given by the interaction coefficient $\Delta \omega = 2 Jk_\mathrm{cross}^2\omega_\mathrm{fy}$. On the other hand, when $ Jk^2\omega_\mathrm{y} \ll \omega_\mathrm{fm}^2 - \omega_\mathrm{H}^2$, the interaction term can be neglected, leading to nearly uncoupled elastic and magnetic dispersion relations. In this regime, the waves are called quasi-elastic or quasi-magnetic waves \cite{Tucker72,Gurevich96}. Hence, the interaction between the elastic and magnetic waves is the strongest when they are (nearly) degenerate, resulting in coupled magnetoelastic waves. By contrast, quasi-noninteracting waves are obtained when their frequencies and/or their wavelengths differ strongly.

The wavenumber at the crossing, $k_\mathrm{cross}$, can be found by equalizing the dispersion relations of the noninteracting systems, \emph{i.e.} $\omega_\mathrm{H}(k_\mathrm{cross}) = \omega_\mathrm{fm}(k_\mathrm{cross})$, and solving for $k_\mathrm{cross}$. For the geometry considered here, the noninteracting dispersion relations are equal when

\begin{equation}
	v_\mathrm{t}k_\mathrm{cross} = (\omega_0+\omega_\mathrm{M} \lambda_{\mathrm{ex}} k_\mathrm{cross}^2)^2 + \omega_\mathrm{M} \left(\omega_0 + \omega_\mathrm{M} \lambda_{\mathrm{ex}} k_\mathrm{cross}^2 + \omega_\mathrm{M} (1-P)P\right) \, ,
\end{equation}

\noindent which needs to be solved iteratively. Note that $P$ is also a function of $k_\mathrm{cross}$ according to Eq. (\ref{eq:P}). Once $k_\mathrm{cross}$ is determined, the interaction coefficient $Jk^2\omega_\mathrm{fy}$ and the gap amplitude can be calculated. In general, the coupling increases strongly for higher wavenumbers $k_\mathrm{cross}$. This originates from the behavior of the magnetostriction and the Villari effect: a shorter wavelength leads to larger gradients of both displacement and magnetization. This increases the magnetoelastic body force (\emph{cf.} Eq. (\ref{eq:force_density1})) and the magnetoelastic field (\emph{cf.} Eq. (\ref{eq:Hmel})), leading to stronger interactions for higher $k_\mathrm{cross}$ values. This behavior opens possibilities for the control of the interaction strength by external parameters. For example, increasing the external applied magnetic field shifts the spin wave dispersion relation to higher frequencies, leading to a larger value of $k_\mathrm{cross}$ and thus a stronger magnetoelastic coupling.

According to the dispersion relation in Eq. (\ref{eq:disp_mel_de2}), two different wave-like solutions exist that correspond to two different magnetoelastic waves. To further describe the characteristics of these waves, the eigenstates need to be calculated. They are given by

\begin{equation}
\label{eq:mel_de_eigenstate}
\begin{bmatrix}
u_\mathrm{t} \\ m_\mathrm{x} \\ m_\mathrm{y}	\end{bmatrix} = N \begin{bmatrix}
1 \\ i \frac{\rho M_\mathrm{s}}{B k} (\omega_{\pm}^2 - \omega_\mathrm{H}^2 ) \\ \frac{\rho M_\mathrm{s} \omega_{\pm}}{Bk \omega_\mathrm{fy}}  (\omega_{\pm}^2 - \omega_\mathrm{H}^2 )
\end{bmatrix} = N \begin{bmatrix}
1 \\ i \frac{\gamma Bk \omega_\mathrm{fy}}{\omega_{\pm}^2 - \omega_\mathrm{fm}^2} \\ \frac{\gamma Bk \omega_{\pm}}{\omega_{\pm}^2 - \omega_\mathrm{fm}^2}
\end{bmatrix}
\end{equation}

\noindent with $N$ being a dimensionless normalization factor. Note that the polarization of the two magnetization components, for both cases $\omega_+$ and $\omega_-$, is clockwise (right-hand) elliptically polarized with ellipticity $|m_\mathrm{x}|/|m_\mathrm{y}| = \omega_\mathrm{fy}/\omega_{\pm}$. The precession described by the $u_\mathrm{t}$ displacement and the $m_\mathrm{x}$ magnetization component is clockwise or counterclockwise (right-hand or left-hand) polarized, depending on the eigenstate $\omega_+$ or $\omega_-$.

Based on the eigenstate, it is possible to determine the variation of the energy associated with the different components of the wave during propagation. There is always a phase difference of $\pi/2$ rad between $m_\mathrm{x}$ and $m_\mathrm{y}$ as well as $u_\mathrm{t}$. This indicates that the energy in the $m_\mathrm{x}$ component is transferred partially to the $m_\mathrm{y}$ and partially to the $u_\mathrm{t}$ components when the waves propagates. Hence, for magnetoelastic waves, there is resonant energy transfer between the elastic and magnetic domain.

\begin{figure}[tb]
\begin{center}
	\includegraphics[width=10cm]{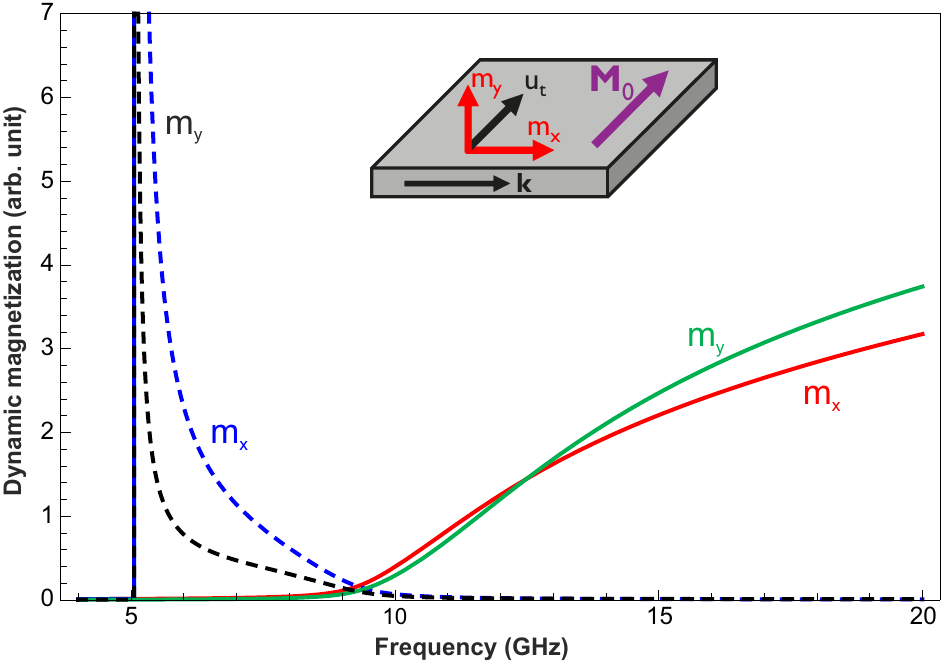}
\end{center}
\caption{Frequency dependence of the dynamic magnetization components of magnetoelastic waves in a 30-nm-thick Ni film. The propagation direction is perpendicular to the magnetization, as shown in the inset. The dashed lines represent the $m_\mathrm{x}$ and $m_\mathrm{y}$ components of the $\omega_+$ state, whereas the solid lines represent the $m_\mathrm{x}$ and $m_\mathrm{y}$ components of the $\omega_-$ state. The external magnetic field is $\mu_0H_\mathrm{ext} = 50$ mT.}
	\label{fig:de_wave_components}    
\end{figure}

The three different regimes identified in the dispersion relation Eq. (\ref{eq:disp_mel_de2}), \emph{i.e.} the quasi-elastic, quasi-magnetic, and magnetoelastic regimes, can also be found in the eigenstates. In the quasi-elastic regime, the dispersion relation approaches the linear dispersion of the elastic wave, \emph{i.e.} $\omega_{\pm}^2 - \omega_\mathrm{H}^2 \approx 0$ and thus $m_\mathrm{x}, m_\mathrm{y} \approx 0$ according to Eq. (\ref{eq:mel_de_eigenstate}). In other words, in the quasi-elastic regime, the total energy is almost completely dominated by the elastic energy \cite{Tucker72,Gurevich96} and the energy transfer to the magnetic system during propagation can be neglected. In the quasi-magnetic regime on the other hand, the dynamic displacement component $u_\mathrm{t}$ is very small and thus the total energy is dominated by the magnetic energy. In the magnetoelastic regime near the anticrossing, the total energy of the wave is distributed between the magnetic and elastic systems. Hence, large parts of the total energy of the wave resonantly oscillates between the magnetic and elastic domains \cite{Tucker72,Gurevich96}. This can also be seen in Fig. \ref{fig:de_wave_components}, which shows the magnetization components for the two branches of the dispersion relation, $\omega_+$ and $\omega_-$, as a function of the frequency. In keeping with the above discussion, the magnetization components have strong amplitudes in the quasi-magnetic and weak amplitudes in the quasi-elastic regime.  

\subsubsection{Wave propagation parallel to the magnetization}

When the propagation direction of the magnetoelastic wave is parallel with the equilibrium magnetization direction, \emph{i.e.} $\theta=0$ in Eq. (\ref{eq:disp_mel_de2}), the magnetic body force acts on both the in-plane and out-of-plane transverse displacement components. Note that in this geometry, the transverse in-plane displacement component is fully aligned along the x-direction, \emph{i.e.} $u_\mathrm{t} = u_\mathrm{x}$. Analogously, both the in-plane and the out-of-plane transverse elastic waves generate magnetoelastic fields that interact with the dynamic magnetization. Hence, both transverse displacement components couple to the backward volume spin waves. Only the longitudinal elastic wave is uncoupled from the magnetic system. Neglecting longitudinal elastic waves, the system of equations in matrix notation then becomes

\begin{equation}
\label{eq:stelsel_mel_bvw}
\begin{bmatrix}
 \omega^2 - \omega_\mathrm{H}^2 &0 & \frac{iBk}{\rho M_\mathrm{s}} & 0 \\
  0 & \omega^2 - \omega_\mathrm{V}^2 &0 & \frac{iBk}{\rho M_\mathrm{s}}  \\
  i\gamma Bk & 0 &  \omega_\mathrm{fx} & -i \omega \\
0 & i\gamma Bk & i \omega & \omega_\mathrm{fy}  
\end{bmatrix} \begin{bmatrix} u_\mathrm{t} \\ u_\mathrm{y} \\ m_\mathrm{x} \\ m_\mathrm{y}
\end{bmatrix} = 0 \,.
\end{equation}

\noindent It is worth noting that both transverse elastic waves have the same dispersion relation and thus $\omega_\mathrm{H}=\omega_\mathrm{V} = v_\mathrm{t}k$ as discussed earlier. 

Again, the homogeneous linear system has only nontrivial solutions when its determinant is zero, which leads to the dispersion relation of the magnetoelastic waves, given by

\begin{equation}
\label{eq:disp_mel_bvw}
	(\omega^2 - \omega_\mathrm{fm}^2)(\omega^2 - \omega_\mathrm{H}^2)(\omega^2 - \omega_\mathrm{V}^2) -Jk^2[\omega_\mathrm{fx}(\omega^2 - \omega_\mathrm{H}^2) + \omega_\mathrm{fy}(\omega^2 - \omega_\mathrm{V}^2)+J k^2]= 0 \,.
\end{equation}

\noindent Three different interaction terms can be identified in this equation. The first interaction term $Jk^2\omega_\mathrm{fx}(\omega^2 - \omega_\mathrm{H}^2)$ represents the interaction between the out-of-plane transverse elastic wave and the backward volume spin wave. The second term $Jk^2\omega_\mathrm{fx}(\omega^2 - \omega_\mathrm{V}^2)$ characterizes the interaction between the in-plane transverse elastic wave and the backward volume spin wave. Thus, these two terms describe an anticrossing near the point where the dispersion relations of noninteracting elastic and magnetic waves would intersect each other. The third interaction term $J^2k^4$ couples all three different waves with each other and thus also generates an interaction between the two transverse elastic waves.

\begin{figure}[tb]
\begin{center}
	\includegraphics[width=10cm]{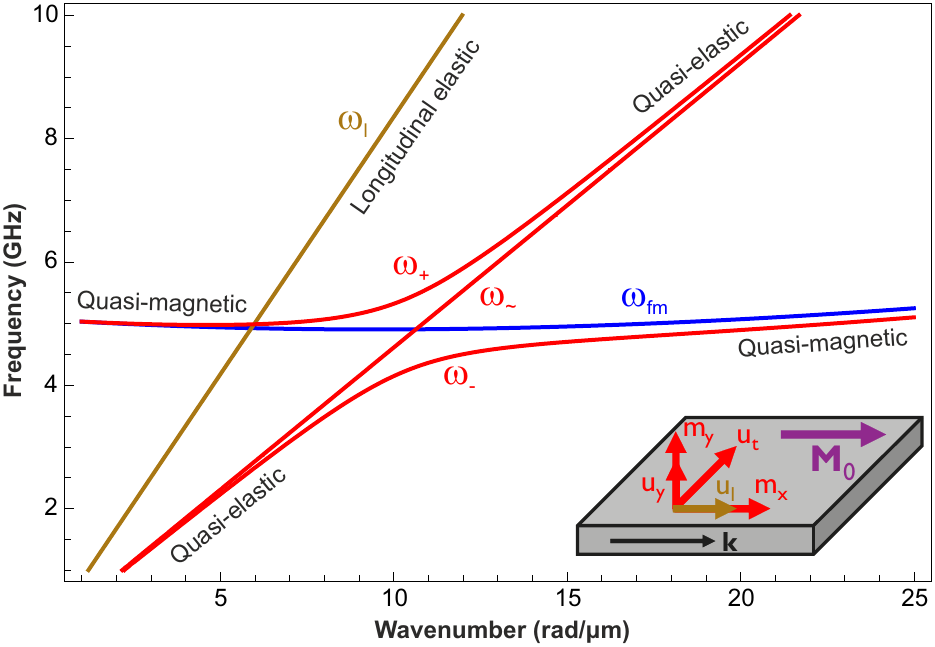}
\end{center}
	\caption{Magnetoelastic wave dispersion relations (red lines) according to Eq. (\ref{eq:disp_mel_bvw}) for a 30-nm-thick Ni film and propagation directions parallel with the magnetization, as shown in the inset. The external magnetic field is $\mu_0H_\mathrm{ext} = 50$ mT. For comparison, the dispersion relations of longitudinal elastic waves (brown line) and uncoupled spin waves (blue line) are also shown.}
	\label{fig:bvw_wave_dispersions}       
\end{figure}

\begin{figure}[tb]
\begin{center}
	\includegraphics[width=10cm]{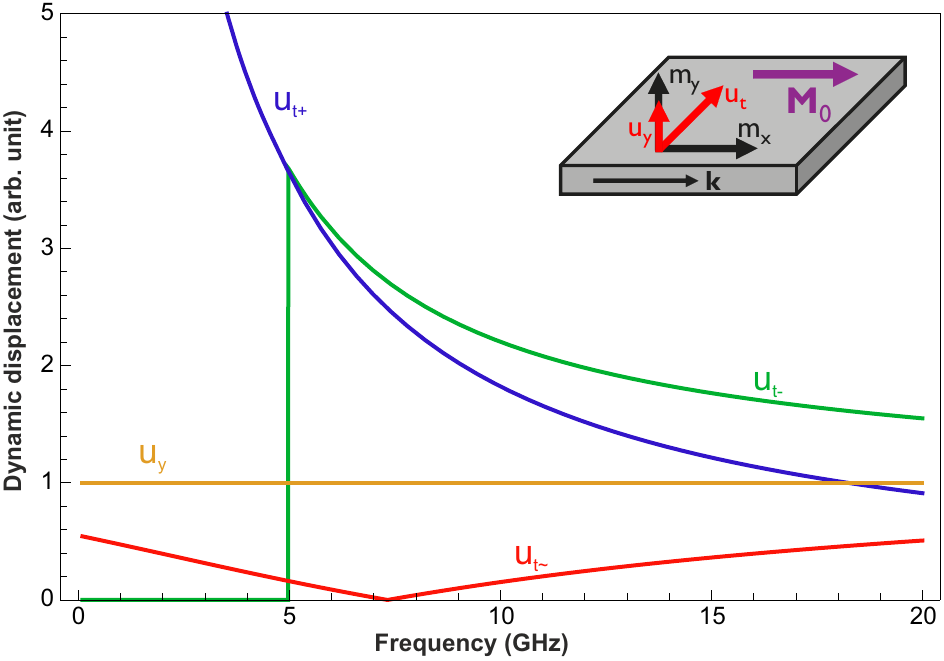}
\end{center}
	\caption{Frequency dependence of the dynamic displacement components for the different magnetoelastic waves in a 30-nm-thick Ni film and an external magnetic field of $\mu_0H_\mathrm{ext} = 50$ mT. The propagation direction is parallel to the magnetization, as shown in the inset. All displacement values are normalized to the out-of-plane component of the displacement $u_\mathrm{y}$ (yellow line). The blue, green and red lines correspond to the in-plane displacement components of the $\omega_+$, $\omega_-$ and $\omega_\sim$ modes, respectively.}
	\label{fig:disp_comp_bvw}       
\end{figure}

\begin{figure}[tb]
\begin{center}
	\includegraphics[width=10cm]{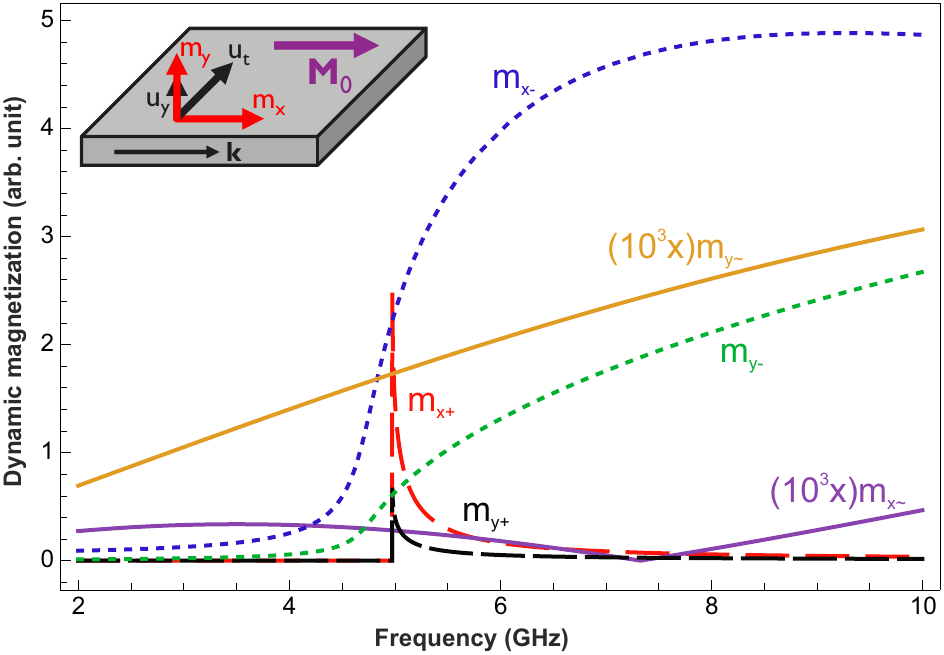}
\end{center}
	\caption{Frequency dependence of the dynamic magnetization components of magnetoelastic waves in a 30-nm-thick Ni film for an external magnetic field of $\mu_0H_\mathrm{ext} = 50$ mT. The propagation direction is parallel to the magnetization, as shown in the inset. The dashed blue and green lines correspond to the $\omega_-$ mode, whereas the dashed red and black line correspond to the $\omega_+$ modes, and the solid lines correspond to the $\omega_\sim$ mode. Note that the magnetization components corresponding to the $\omega_{\sim}$ mode are multiplied by a factor of $10^3$.}
	\label{fig:mag_comp_bvw}       
\end{figure}

Figure \ref{fig:bvw_wave_dispersions} shows the different dispersion relations for material parameters corresponding to Ni, as mentioned above. To better understand their behavior, the corresponding eigenstates of the different magnetoelastic waves are calculated. The eigenstates are given as a function of the angular frequency of the magnetoelastic wave, $\omega$, by 

\begin{equation}
\label{eq:eig_mel_bvw}
\begin{bmatrix}
u_\mathrm{t} \\ u_\mathrm{y} \\ m_\mathrm{x} \\ m_\mathrm{y}	\end{bmatrix} = N \begin{bmatrix}
\frac{-i}{\omega (\omega^2 - \omega_\mathrm{H}^2)} (\omega_\mathrm{fy} (\omega^2 - \omega_\mathrm{V}^2) + Jk^2) \\
1 \\
- \frac{\rho M_\mathrm{s} }{Bk \omega} (\omega_\mathrm{fy} (\omega^2 - \omega_\mathrm{V}^2) + Jk^2) \\
i \frac{\rho M_\mathrm{s} }{Bk} (\omega^2 - \omega_\mathrm{V}^2)
\end{bmatrix} 
\end{equation}

\noindent with $N$ being a dimensionless normalization constant. The different displacement components of the magnetoelastic waves are plotted in Fig. \ref{fig:disp_comp_bvw} as a function of the frequency, whereas Fig. \ref{fig:mag_comp_bvw} shows the different magnetization components. As above, Ni material parameters were assumed, and the external magnetic field was $\mu_0H_\mathrm{ext} = 50$ mT.

In the following, the different eigenstates and their properties are discussed. The upper $\omega_\mathrm{+}$ and lower $\omega_\mathrm{-}$ branches of the dispersion relation both correspond to clockwise (right-hand) elliptically polarized waves for the magnetization and displacement, \emph{i.e.} $m_\mathrm{y}/m_\mathrm{x} = i|m_\mathrm{y}|/|m_\mathrm{x}|$ and $u_\mathrm{y}/u_\mathrm{x} = i |u_\mathrm{y}|/|u_\mathrm{x}|$ \cite{Tucker72,Gurevich96}. In both cases, the in-plane magnetization component is always larger than the out-of-plane component, \emph{i.e.} $m_\mathrm{x} > m_\mathrm{y}$, since the demagnetization field is strongest in the out-of-plane direction. Concerning the displacement components, the two branches behave differently. For the $\omega_\mathrm{+}$ eigenstate, the in-plane and out-of-plane displacement components have the same order of magnitude at GHz frequencies. On the other hand, for the $\omega_\mathrm{-}$ state, $u_\mathrm{t}$ is dominant at low frequencies, whereas $u_\mathrm{y}$ becomes dominant at high frequencies. This behavior can also be seen in Fig. \ref{fig:disp_comp_bvw}.

The dispersion relation corresponding to the third magnetoelastic eigenstate is also shown in Fig. \ref{fig:bvw_wave_dispersions}, labelled $\omega_{\sim}$. The dispersion is nearly linear and falls slightly below the dispersion relation for uncoupled transverse elastic waves, which was discussed in Section \ref{sec:elastic waves} \cite{Tucker72,Gurevich96}. The magnetization and the displacement components corresponding to this state are both counterclockwise (left-hand) elliptically polarized. For uncoupled backward volume spin waves, counterclockwise polarization corresponds to evanescent spin waves. However, such evanescent spin waves can still couple to left-hand polarized displacement waves, resulting in left-hand polarized \emph{propagating} magnetoelastic waves. Nevertheless, the magnetization components for this magnetoelastic mode remain very weak. This is also seen from Fig. \ref{fig:mag_comp_bvw} where the magnetization components corresponding to the $\omega_\sim$ branch have three orders of magnitude lower amplitude than the magnetization components of the $\omega_+$ and $\omega_-$ branches. In terms of displacement, the $u_\mathrm{y}$ displacement component is dominant for the $\omega_{\sim}$ mode for a wide frequency range. This is also illustrated in Figs. \ref{fig:disp_comp_bvw} and \ref{fig:mag_comp_bvw}.

Because of the coupling to the spin wave system, magnetoelastic waves show some peculiarities in the quasi-elastic regime, where the wave energy is largely dominated by the elastic energy. As shown above, the $J^2k^4$ interaction term couples all waves with each other. Consequently, the two transverse elastic waves become also coupled. For interacting waves, it is impossible to share the same frequency--wavenumber couple, \emph{i.e.} it is impossible to have degenerate points in the dispersion relations. As a result, the two quasi-elastic branches do not overlap anymore which is in contrast to their original behavior without magnetoelastic interactions (see Section \ref{sec:elastic waves}). Therefore, at all frequencies, a small wavenumber shift remains present between the two quasi-elastic branches, even in the quasi-elastic regime where the displacement components are large and magnetization components are weak. Hence, even though almost all the wave energy is in the elastic system, the interaction between the two transverse displacement components is mediated by the magnetic system, leading to an indirect coupling of the two elastic waves via the magnetic system. This interaction is proportional to $k^4$ and thus strongly depends on the wavelength. 

Moreover, the polarization of the displacement in the quasi-elastic regime also shows a peculiar behavior. One of the two waves in the quasi-elastic regime correspond to a clockwise (right-hand) polarized wave and the other to a counterclockwise (left-hand) polarized wave, as discussed above. Hence, excitation at a single angular frequency $\omega$ in the quasi-elastic regime leads to two different magnetoelastic waves with different wavelength and opposite polarization. Their amplitudes in function of time and space can be written as

\begin{equation}
	\mathbf{u}_+ = \begin{bmatrix}
	|u_\mathrm{t+}| \\ i |u_\mathrm{y+}|
	\end{bmatrix} e^{i\omega t + k_+ z}\, \, \, \, \text{and} \, \,\,\, \mathbf{u}_- = \begin{bmatrix}
	|u_\mathrm{t-}| \\ -i|u_\mathrm{y-}|
	\end{bmatrix} e^{i\omega t + k_- z} 
\end{equation}

\noindent with $u_\mathrm{t+}=u_\mathrm{t}(k_+)$ and $u_\mathrm{t-}=u_\mathrm{t}(k_-)$ given by Eq. (\ref{eq:eig_mel_bvw}). The total wave is the sum of both individual waves

\begin{equation}
	\mathbf{u}_\mathrm{tot} = \begin{bmatrix}
	|u_\mathrm{t+}|e^{ik_+ z}   + |u_\mathrm{t-}|e^{ik_- z}     \\  i  \left(|u_\mathrm{y+}|e^{ik_+ z}  - |u_\mathrm{y-}|e^{ik_- z}   \right)  
	\end{bmatrix}  e^{i\omega t} \,.
\end{equation}

\noindent The difference in amplitude between the clockwise and counterclockwise polarized components results in an elliptical polarization of the total displacement. The different wavenumbers of the two individual waves ($k_+$ and $k_-$) result in the rotation of the major and minor axes of the ellipsoid described the tip of the displacement vectors during wave propagation \cite{Matthews62,Vlasov60}. This is similar to the Faraday effect for electromagnetic waves and is also called acoustic wave rotation. 

The dispersion relation of backward volume spin waves is rather flat in the dipolar--exchange regime, leading to an interesting property of magnetoelastic waves in this geometry. As shown in Fig. \ref{fig:bvw_wave_dispersions} at frequencies around 4--5 GHz, the magnetoelastic coupling leads to the formation of a pseudobandgap for clockwise (right-hand) polarized elastic waves at the anticrossing. On the other hand, due to the flatness of the dispersion relation, counterclockwise (left-hand) polarized magnetoelastic waves can still exist in this pseudobandgap. Hence, in this frequency range, only pure magnetoelastic waves or quasi-magnetic waves with weak displacement components can be excited. This pseudobandgap formation is a general result when waves with a rather flat dispersion relation interact with waves with a steep dispersion relation near the crossing point.

\subsubsection{Arbitrary propagation direction}

We now consider an arbitrary propagation direction of the magnetoelastic wave with respect to the equilibrium magnetization. In this case, the magnetoelastic body force interacts with all displacement components. Conversely, all displacement components generate magnetic fields, which interact with the magnetization. Hence, all magnetization and displacement components become coupled with each other. Again, nontrivial wave-like solutions only exist when the determinant of the linear system in Eq. (\ref{eq:system_of_melwaves2}) is zero, which leads to the dispersion relation

\begin{equation}
\label{eq:general_dispersion}
\begin{gathered}
(\omega^2 - \omega_l^2)[(\omega^2 - \omega_\mathrm{t}^2)^2(\omega^2-\omega_\mathrm{fm}^2) \\
- (\omega^2-\omega_\mathrm{t}^2) Jk^2(\omega_\mathrm{fx}\cos^2(\theta) + \omega_\mathrm{fy}\cos^2(2\theta)) - J^2 k^4 \cos^2(2\theta)\cos^2(\theta)] \\
-(\omega^2-\omega_\mathrm{t}^2)Jk^2[\omega_\mathrm{fy}(\omega^2-\omega_\mathrm{t}^2)\sin^2(2\theta) + Jk^2 \sin^2(2\theta)\cos^2(\theta)]=0 \,.
\end{gathered}
\end{equation}

\noindent Note that for $\theta=\pi/4$, the coupling between the magnetic and the longitudinal elastic wave reaches a maximum, whereas for $\theta=0$ and $\theta=\pi/2$, the coupling is zero.

\begin{figure}[tb]
\begin{center}
	\includegraphics[width=10cm]{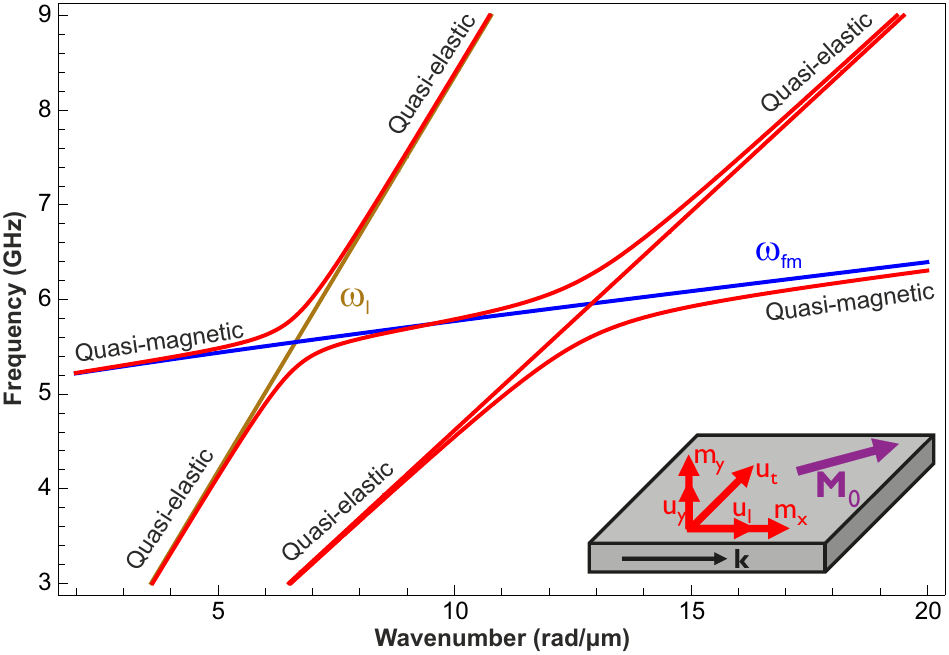}
\end{center}
	\caption{Magnetoelastic wave dispersion relations (red lines) according to Eq. (\ref{eq:general_dispersion}) for a 30-nm-thick Ni film nm and an angle of 30$^\circ$ between the propagation direction and the magnetization. The external field is $\mu_0H_\mathrm{ext} = 50$ mT. For comparison, the dispersion relations of longitudinal elastic waves (brown line) and uncoupled spin waves (blue line) are also shown.}
	\label{fig:int_wave_dispersions}      
\end{figure}

The dispersion relations of the resulting magnetoelastic waves are plotted in Fig. \ref{fig:int_wave_dispersions} for material parameters of Ni and $\theta=\pi/6$. For each frequency, multiple magnetoelastic waves exist with different wavelengths. Since the system of equations was reduced to a set of linear differential equations by assuming weak dynamic components, every linear combination of these different magnetoelastic waves is also a solution of the system. The waves can be equally excited via dynamic magnetic fields or mechanical forces. Therefore, it is possible to generate elastodynamics via the magnetization or, \emph{vice versa}, magnetization dynamics via the displacement in magnetostrictive materials. 

The total energy of a magnetoelastic wave consists of several contributions. The magnetic energy contribution is determined by the dynamic components $m_\mathrm{x}$ and $m_\mathrm{y}$. In this paper, only Zeeman, dipolar, and exchange energy interactions were considered, although other magnetic interactions, such as the magnetocrystalline \cite{Dreher12,Thevenard14,Gowtham15} or the Dzyaloshinskii–Moriya interaction \cite{Verba18} may also contribute to the total energy. The magnetic energy is complemented by the energy of the elastic waves, which consists of both elastic and kinetic energy contributions, and is fully determined by the displacement components and their time derivatives, given in Eqs. (\ref{eq:elastic_energy}) and (\ref{eq:kinetic_energy}), respectively. A third energy contribution stems from the magnetoelastic interaction, as described by Eq. (\ref{eq:Emel}). Just as spin waves (\emph{cf.} Eq. (\ref{eq:electric_field})), magnetoelastic waves also comprise an electric field component. However, at GHz frequencies, the magnetostatic approximation is typically valid and therefore the energy contribution of the electric field is small and can typically be neglected. Nonetheless, this ceases to be accurate when frequencies approach the THz range where the magnetostatic approximation no longer holds.

During the propagation of the magnetoelastic wave, the energy oscillates between the different energy contributions. For strongly interacting waves near the anticrossing point, a large part of the energy oscillates resonantly between the elastic and magnetic domains. This energy transfer is characterized by a specific energy transfer length, which characterized the distance that is necessary to transfer the energy from one system to the other \cite{Graczyk17}. By contrast, in the quasi-elastic regime most of the energy remains in the elastic system during propagation, whereas in the quasi-magnetic regime most energy remains in the magnetic system \cite{Tucker72,Gurevich96}. 

In this paper, it was assumed that the wavelength of the magnetoelastic wave is much larger than the thickness of the film. In this case, the dynamic magnetization and displacement are approximately uniform over the film thickness. However, if the thickness becomes comparable to the wavelength, different thickness modes can arise. In the magnetic domain, these are called perpendicularly standing spin waves and in the elasticity domain, these are called Lamb waves. The magnetoelastic coupling of such waves is beyond the scope of this paper and will generally require numerical calculations.

\subsection{Damping of magnetoelastic waves}

So far, all waves have been considered to be lossless and their damping was neglected. However, in real systems, magnetoelastic waves are expected to decay during propagation. Since their decay length is of great practical interest, we will present in this last part a brief introduction on the damping of magnetoelastic waves. More detailed discussions can be found in \cite{Rezende69,Widom10,Gurevich15,Rossi05,Streib18}. 

Several different energy loss mechanisms exist, which dampen the dynamics of the magnetization and the displacement. In the semi-classical continuum theory used in this paper, it is common to subsume all the different loss mechanisms in a single phenomenological damping term, which is then included in the equation of motion. The damping for the magnetization dynamics is captured by the damping term in the LLG equation, characterized by the phenomenological Gilbert damping parameter $\alpha$. Analogously, for elastic waves, damping can be introduced into the equations of motion via phenomenological complex stiffness constants. 

The addition of the damping terms to the equations of motion results in energy dissipation of the dynamic system. As a consequence, the amplitude of the plane waves considered above now decays in time and space. Therefore, the plane wave ansatz to solve the equations of motion needs to be modified by adding an exponential decay factor. As a result, the frequency of the wave becomes complex, \emph{i.e.}
\begin{equation}
\mathbf{w}(\mathbf{r},t) e^{i ((\omega_r +i \omega_i) t + \mathbf{k}\cdot \mathbf{r})} = \mathbf{w}(\mathbf{r},t) e^{-t/\tau} e^{i (\omega_r t + \mathbf{k}\cdot \mathbf{r})} = \mathbf{w}(\mathbf{r},t) e^{-x/\delta} e^{i (\omega_r t + \mathbf{k}\cdot \mathbf{r})}
\end{equation}

\noindent with $\tau=1/\omega_i$ the lifetime, $\delta=v_g \tau$ the mean free path, and $\mathbf{w}(\mathbf{r},t) = [u_\mathrm{x},u_\mathrm{y},u_\mathrm{z},m_\mathrm{x},m_\mathrm{y}]^T$ the dynamic components of the wave. Note that the lifetime characterizes the decay of the wave in time and the mean free path characterizes the attenuation of the wave in space. 

To determine the decay characteristics of a wave, the imaginary part of its frequency needs to be assessed. This can be achieved within the approach described above, which is based on finding nontrivial solutions of homogeneous linear systems by calculating the roots of their determinants. The real part of the resulting frequency still represents the dispersion relation, whereas the imaginary part originates from the additional damping terms and represents the inverse of the lifetime. 

For spin waves in a ferromagnetic medium, the lifetime is found by solving the LLG equation and is given by

\begin{equation}
	\tau_\mathrm{fm} =  \frac{2}{\alpha (\omega_\mathrm{fx}+\omega_\mathrm{fy})} \,.
\end{equation}

\noindent The lifetime at GHz frequencies is typically in the order of ns in metallic ferromagnets, such as Ni considered in this paper, and in the order of $\mu$s for low-damping magnetic insulators such as Yttrium Iron Garnet (YIG). On the other hand, much less is know for elastic waves at GHz frequencies although estimates suggest that the lifetime is similar to that of spin waves. Experimentally, it is typically found that the mean free path of (surface) elastic waves at these frequencies is somewhat larger than these of spin waves \cite{Dreher12,Verba18,Graczyk17,Li17}, although the topic still requires further research.

In the case of magnetoelastic waves, analytical derivations of the lifetimes and decay lengths are rather complex. In the quasi-elastic regime, it is clear that the lifetime is strongly determined by the lifetime of the elastic wave. The energy of quasi-elastic waves is almost completely stored in the elastic system, with only a negligible part in the magnetic system, and thus the dissipation due to the magnetic loss should have no influence on the overall dissipation. An analogous argument can be made for the quasi-magnetic regime, where magnetic properties and lifetimes should determine the decay of the magnetoelastic waves. 

In the strongly coupled magnetoelastic regime, \emph{i.e.} near the anticrossing, no simple conclusion can be drawn. In this regime, the energy is distributed between magnetic and elastic domains and is transferred forth and back during propagation. Therefore both magnetic and elastic losses contribute to the total energy dissipation. One may expect in such a case that the lifetime of a magnetoelastic wave is given by a suitable weighted average of the lifetimes of magnetic and elastic waves. In general, it depends on multiple parameters, such as the orientation of the static magnetization, the interaction coefficient, the wavenumber, \emph{etc}. Further work is required to fully understand in particular the effect of the magnetoelastic interactions on the lifetime of strongly coupled magnetoelastic waves. By contrast, the group velocities of magnetoelastic waves are well understood and can be calculated from the dispersion relations, so the assessment of man free paths is straighforward if the lifetime is known.

\begin{acknowledgments}

This work has been supported by imec's industrial affiliate program on beyond-CMOS logic as well as by the European Union's Horizon 2020 research and innovation program within the FET-OPEN project CHIRON under grant agreement No. 801055. F.V. acknowledges financial support from the Research Foundation -- Flanders (FWO) through grant No. 1S05719N.
\end{acknowledgments}

%
%

\end{document}